\RequirePackage{booktabs}

\documentclass{sn-jnl}

\usepackage{natbib}
\usepackage{graphicx}
\usepackage{multirow}
\usepackage{amsmath,amssymb,amsfonts}
\usepackage{amsthm}
\usepackage{mathrsfs}
\usepackage[title]{appendix}
\usepackage{textcomp}
\usepackage{manyfoot}
\usepackage{booktabs}
\usepackage{bbm}
\newcommand{\dd}{\mathrm{d}}
\newcommand{\e}{\mathrm{e}}
\newcommand{\Ord}{\mathrm{O}}
\newcommand{\one}{\mathbbm{1}}
\newcommand{\Beta}{\mathrm{B}}
\newcommand{\av}[1]{\langle#1\rangle}

\begin{document}

\title{Fast sampling and model selection for Bayesian mixture models}

\author*{\fnm{M. E. J.} \sur{Newman}}

\affil{\orgdiv{Center for the Study of Complex Systems}, \orgname{University of Michigan},\\ \city{Ann Arbor}, \state{Michigan} \postcode{48109}, \country{USA}}

\abstract{We study Bayesian estimation of mixture models and argue in favor of fitting the marginal posterior distribution over component assignments directly, rather than Gibbs sampling from the joint posterior on components and parameters as is commonly done.  Some previous authors have found the former approach to have slow mixing, but we show that, implemented correctly, it can achieve excellent performance.  In particular, we describe a new Monte Carlo algorithm for sampling from the marginal posterior of a general integrable mixture that makes use of rejection-free sampling from the prior over component assignments to achieve excellent mixing times in typical applications, outperforming standard Gibbs sampling, in some cases by a wide margin.  We demonstrate the approach with a selection of applications to Gaussian, Poisson, and categorical models.}

\maketitle

\section{Introduction}
Mixture models provide a powerful framework for model-based clustering, in which data are partitioned into clusters or components, each with their own characteristic distribution, and model fits return an estimate of both the partition and the individual distributions \citep{TSM85,MP00,MMR05,GMR22}.  The fit is often performed using an expectation-maximization (EM) algorithm that returns maximum likelihood or maximum a posteriori (MAP) estimates of model parameters and a complete posterior distribution over the assignment of data to components \citep{DLR77,MP00,Gelman13,MLR19}.  EM algorithms however have some shortcomings.  In addition to returning only point estimates of the parameters, they can lead to overfitting of the data or underdetermination of the parameters when the number of parameters is large or the data are sparse, and they provide no direct way of estimating the number of components---commonly, one just performs repeated fits with different numbers of components, then selects among them using, for example, likelihood ratios or the Bayesian information criterion, but this a costly and cumbersome procedure, making it tempting to neglect this important step of the analysis \citep{Everitt88,LD97,BS03b,RD06,NAM07,Hoijtink10,MR14,NVDL17}.

Alternatively, we can take a Bayesian approach and impose a prior on the parameters then integrate them out of the model to yield a marginal posterior over the component assignments alone.  This obviates any problems with point estimates of the parameters, but it has its own shortcomings, at least as typically implemented.  The integration is most commonly approximated numerically by Gibbs sampling \citep{BCRR97,Gelfand00,Neal00,Fruhwirth06,GMR22}, which can be computationally efficient but again does not provide a direct route for estimating the number of components: the number cannot itself be sampled because the size of the parameter space is not constant when the number of components is varying.  Reversible jump Monte Carlo and similar methods \citep{Green95,PS96,RG97,Stephens00b} can get around this issue, although they are more computationally demanding and have been described as ``notoriously hard to tune and may lead to poor mixing'' \citep{BGGG25}.  More often, one again just performs repeated analyses with various (fixed) numbers of components and selects among the resulting fits, but this can also be numerically costly, since it requires repeated Monte Carlo runs, with most of the data being discarded in the final analysis.

An alternative is to marginalize analytically over the model parameters, which can be done in closed form for many of the most common mixture models, using suitable priors \citep{Nobile04,SRE06,BCG10,NR16,WWM16}.  Then one can sample directly from the integrated posterior using one of several proposed sampling schemes, such as the allocation sampler of \citet{NF07} or the modified Gibbs samplers of \citet{JN04} and \citet{Porteous08}.  This approach is not widely used---for example, the otherwise comprehensive review of~\cite{MR14} fails to mention it even in passing---but it is potentially promising for practical, robust mixture modeling because it again removes any issues with point estimates and it allows one to sample not only the component assignments but also the number of components, and hence estimate both at the same time without the need for a separate model selection procedure.  Some authors have suggested that it may suffer from slow mixing \citep{NF07,MH18}, but we argue that, when implemented correctly, it can be highly numerically efficient.  It is a version of this approach that we study in this paper.

Standard mixture models, as commonly formulated, also suffer from a technical, but important, difficulty: the existence of empty components \citep{RM11}.  In many models the number of observations in a component can be zero, and while this is arguably acceptable for a model with a fixed number of components, when the number of components is a free random variable it generates ambiguity, because a given division of observations into components can be represented in more than one way in the model.  For instance, we could divide observations into two components, or we could divide them into three components, one of which is empty.  This in turn creates difficulties when estimating the number of components---do we have two components or do we have three?  (We note that some authors use the word ``cluster'' to refer to a non-empty component, although others use the words cluster and component interchangeably.  In this paper we will specifically use the phrase ``non-empty component'' for the sake of clarity.)

One solution to the problem of empty components is to employ a sparse model with a prior that penalizes empty components and then count only non-empty ones \citep{VWRM15,FM19}.  Another is to use an infinite model with a Dirichlet-process prior and again count only non-empty components \citep{Neal00,HHMW10}.  Here we argue for a third approach that straddles the boundary between infinite mixtures and sparse finite ones.

Our contributions are two-fold.  First, we advocate in favor of a variant prior on the component assignments, closely similar to a traditional Dirichlet-categorical prior, but differing in that it forbids the existence of empty components.  Second, and more importantly, we describe a Monte Carlo algorithm for sampling from the resulting integrated posterior that exploits the structure of the prior to create a more efficient algorithm that significantly outperforms other state-of-the-art methods.  Our approach can in principle be applied to any mixture model for which a closed-form expression exists for the integrated posterior.  Models falling in this category include many of the most widely used models, such as univariate and multivariate Gaussian mixtures with normal/gamma priors (normal distribution over the mean and gamma distribution over the inverse variance or precision); Gaussian mixtures with normal/inverse-gamma priors (inverse-gamma distribution over the variance); Dirichlet-process Gaussian models with normal/Wishart priors or normal/inverse-Wishart priors; Poisson mixtures with gamma priors; geometric mixtures with gamma priors; multivariate Bernoulli and categorical models; and binomial or multinomial mixtures with beta or Dirichlet priors.

Crucially, the algorithm we describe does not fix the number of components, but allows it to vary freely, being sampled along with the component assignments themselves.  This allows one trivially to determine the posterior distribution over the number of components and hence to infer the number of components as an integral part of the calculation.  No additional model selection procedure, for instance using the Bayesian information criterion, is necessary, and only a single run of the algorithm is needed to determine both the number of components and the assignment of observations to components.  We show that this approach is consistently able to recover the correct number of components in synthetic tests.

As examples of our methods, we apply them to Gaussian, Poisson, and categorical mixture models on a range of example data sets.  We provide implementations in the C and Python programming languages that can perform millions of Monte Carlo steps per second on standard hardware, sufficient that calculations on typical data sets can be completed in a few seconds, and we give timing results showing that this is substantially faster than competing Monte Carlo methods.

\section{Bayesian mixture models}
\label{sec:model}
Consider a data set of $N$ observations $x_i$ with $i=1\ldots N$, where an observation can be a single number or categorical variable, or a (possibly heterogeneous) collection of such variables.  We assume the data are drawn from a mixture model defined as follows.

First, we pick the number~$k$ of components, which can take any value between~1 (all observations in a single component) and~$N$ (every observation the sole member of its own component).  For instance, one could use a uniform (uninformative) prior on~$k$, which means
\begin{equation}
P(k) = {1\over N}.
\label{eq:pk}
\end{equation}
Other choices are also possible: if one had a particular belief about the expected value of~$k$, for instance, then the maximum-entropy (least informative) prior would be a geometric distribution $P(k) \propto a^k$ for some $a<1$.  In our example applications we use a uniform prior on~$k$, but we derive the theory for general~$P(k)$.  In this respect our approach offers more flexibility than, for example, models with a Dirichlet process prior, which imposes a specific choice of~$P(k)$.

Given~$k$, each observation~$i$ is assigned to a component $z_i = 1\ldots k$.  In the first instance, let us suppose the components are drawn from a $k$-dimensional symmetric Dirichlet-categorical distribution with some concentration parameter~$\alpha$:
\begin{equation}
z \sim \text{DirichletCategorical}(N,k,\alpha).
\end{equation}
That is, we choose a set of $k$ probabilities~$\pi_r$ with $r=1\ldots k$ from the symmetric Dirichlet distribution
\begin{equation}
P(\pi|k,\alpha) = {1\over\Beta(\alpha,\alpha,\ldots)}
  \prod_{r=1}^k \pi_r^{\alpha-1},
\end{equation}
where
\begin{equation}
\Beta(x_1,x_2,\ldots) = {\prod_r \Gamma(x_r)\over\Gamma\bigl(\sum_r x_r\bigr)}
\end{equation}
is the multivariate beta function and $\pi$ is shorthand for the entire set of probabilities~$\pi_r$.  Now we choose each $z_i$ independently from the categorical distribution with probabilities~$\pi_r$, so that the probability of a particular assignment of components for all $N$ observations is
\begin{equation}
P(z|\pi) = \prod_{i=1}^N \pi_{z_i} = \prod_{r=1}^k \pi_r^{n_r},
\end{equation}
where $n_r$ is the number of observations in component~$r$.  Marginalizing over the probabilities, we then get
\begin{align}
P(z|k,\alpha) &= \int P(\pi|k,\alpha) P(z|\pi) \>\dd\pi
   = {1\over\Beta(\alpha,\alpha,\ldots)} \int \prod_{r=1}^k \pi_r^{n_r+\alpha-1} \>\dd\pi
     \nonumber\\
  &= {\Beta(n_1+\alpha,n_2+\alpha,\ldots)\over\Beta(\alpha,\alpha,\ldots)}
   = {\Gamma(k\alpha)\over\Gamma(N+k\alpha)}
                \prod_{r=1}^k {\Gamma(n_r+\alpha)\over\Gamma(\alpha)}.
\label{eq:empty}
\end{align}

Given these component assignments, we now generate the values of the observations~$x_i$ themselves.  Before getting to that, however, we point out a shortcoming of the distribution in Eq.~\eqref{eq:empty}, that it can result in empty components, as mentioned in the introduction.  There is nothing to stop $n_r$ from being zero, and indeed correct normalization requires that it must be zero some portion of the time, but it is unclear what this means.  What does it mean to have three components, one of which is empty?  How is that different from two non-empty components?  Unless the value of $k$ is known \textit{a~priori}, which we here assume it is not, there is no practical difference between these situations, and yet they appear as distinct component assignments in the model \citep{RM11,FM19}.

These issues can be disposed of, however, by an easy modification.  We simply forbid empty components, meaning we consider only component assignments where each component contains at least one observation.   We can achieve this by a generative process in which we first assign one observation, chosen uniformly at random, to each of the $k$ components, then assign the remainder following a Dirichlet-categorical distribution with concentration parameter~$\alpha$.  There are $N!/(N-k)!$ ways to choose the initial~$k$ assignments and probability $\prod_r \pi_r^{n_r-1}$ of a particular assignment of the remaining $n_r-1$ members of each component.  Because any of the members of a component could be chosen as the first member, there are also $\prod_r n_r$ ways to generate the same complete set of assignments, so the total probability of any particular assignment~is
\begin{equation}
P(z|\pi) = {(N-k)!\over N!} \prod_{r=1}^k n_r \pi_r^{n_r-1},
\end{equation}
and
\begin{align}
P(z|k,\alpha) &= \int P(\pi|k,\alpha) P(z|\pi) \>\dd\pi
   = {1\over\Beta(\alpha,\alpha,\ldots)} {(N-k)!\over N!} 
     \prod_{r=1}^k n_r \int \prod_{r=1}^k \pi_r^{n_r+\alpha-2} \>\dd\pi
     \nonumber\\
  &= {(N-k)!\over N!} \Biggl[ \prod_{r=1}^k n_r \Biggr]
     {\Beta(n_1+\alpha-1,n_2+\alpha-1,\ldots)\over\Beta(\alpha,\alpha,\ldots)}
     \nonumber\\
  &= {(N-k)!\over N!} {\Gamma(k\alpha)\over\Gamma(N+k(\alpha-1))}
     \prod_{r=1}^k n_r {\Gamma(n_r+\alpha-1)\over\Gamma(\alpha)} \nonumber\\
  &= {1\over N!} (N-k)\,\Beta(N-k,k\alpha)
     \prod_{r=1}^k n_r {\Gamma(n_r+\alpha-1)\over\Gamma(\alpha)}.
\label{eq:nonempty}
\end{align}
This is the prior we use in this paper.  Our methods can be applied also to the (more conventional) form in Eq.~\eqref{eq:empty}, but we consider~\eqref{eq:nonempty} a more satisfactory version of the model from both a theoretical and practical point of view.

Our approach can be applied for any value of the concentration parameter~$\alpha$, but the most common choice is $\alpha=1$, which corresponds to a uniform distribution over the Dirichlet probabilities~$\pi_r$ and hence also over the sizes~$n_r$ of the components.  For this~$\alpha$, Eq.~\eqref{eq:nonempty} simplifies to
\begin{equation}
P(z|k) = {N-1\choose k-1}^{\!-1} {\prod_r n_r!\over N!}.
\label{eq:pzk}
\end{equation}
This is the choice we make for the example calculations in this paper, although our algorithms work for any value of~$\alpha$.

Once components have been assigned, the data~$x_i$ are drawn independently from some distribution $P(x_i|\theta_r)$ that depends on parameter values~$\theta_r$ that are unique to the component~$r$.  (The single symbol $\theta_r$ may represent multiple parameters.)  In a Gaussian model, for instance, $\theta_r$~might describe the mean and variance of the Gaussian distribution of data in component~$r$.  Then the complete data likelihood is
\begin{equation}
P(x|k,z,\theta) = \prod_i P(x_i|\theta_{z_i}).
\label{eq:likelihood}
\end{equation}
Assuming a prior~$P(\theta|k)$ on~$\theta$, the joint posterior on $k$, $z$, and the parameters is then
\begin{equation}
P(k,z,\theta|x) = {P(x|k,z,\theta)P(z|k)P(\theta|k)P(k)\over P(x)}.
\end{equation}

\subsection{Fitting the model}
At this point, one commonly proceeds in one of two ways.  The first is to fit the model to the data using an EM algorithm, but for the reasons outlined in the introduction we prefer a Bayesian approach.  The most common Bayesian method uses a Gibbs sampler which, for given $x$ and~$k$, samples alternately from the marginal distributions of the parameters~$\theta_r$ and the component assignments~$z_i$ \citep{Gelfand00,Neal00,GMR22}.  Discarding the~$\theta_r$ then gives us a true sample from the marginal posterior~$P(z|x,k)$, which can be used to estimate component assignments.  This method does not however give us a direct way of estimating~$k$---typically $k$ is instead determined by selecting among models with various fixed values of~$k$ using, for instance, the Bayesian information criterion \citep{RD06,MR14}.  This adds extra work and complexity to the calculation.  Extensions of the Gibbs sampling approach have been proposed that allow for direct estimation of~$k$, such as reversible jump Monte Carlo \citep{Green95,RG97}, but here we take a different approach.  We integrate the parameters~$\theta$ out of the likelihood exactly and then sample directly from the resulting marginal distribution.  We write
\begin{equation}
P(k,z|x) = {P(x|k,z)P(z|k)P(k)\over P(x)},
\label{eq:posterior}
\end{equation}
where
\begin{equation}
P(x|k,z) = \int P(x|k,z,\theta) P(\theta|k) \>d\theta.
\label{eq:marginal}
\end{equation}
With an appropriate prior~$P(\theta|k)$ the integral can be completed in closed form for many of the most common mixture models.  This approach allows us, in principle at least, to make estimates of $k$ as well as the component assignments~$z$, because $k$~is now a free parameter of the distribution and we are at liberty to estimate it just as we would any other parameter, for instance by sampling its value.  In particular, the dimension of the parameter space does not vary with~$k$, as it does in the model before integration, so no variable-dimension sampling method, such as reversible jump Monte Carlo, is required.

In practice, however, implementing this approach is not trivial.  One might imagine that one could sample directly from~\eqref{eq:posterior} using Metropolis-Hastings Monte Carlo for instance, but, while this can be done, it typically has low numerical efficiency.  A~better approach is to use a so-called collapsed Gibbs sampler, although most such samplers do not sample values of~$k$ and hence a separate model selection step is still needed.  Exceptions include the ``split-merge'' method of \cite{JN04} and the allocation sampler of \citet{NF07} for Gaussian mixture models and its extension to other models \citep{WWM16}.  These methods do sample~$k$, although the former is computationally demanding and the latter relies on a Metropolis-Hastings style rejection algorithm which again has low efficiency.

In this paper we describe a method for sampling directly and efficiently from the marginal posterior of Eq.~\eqref{eq:posterior}.  Our approach employs a combination of techniques to create an algorithm that is both simple to implement and gives results comparable to previous approaches in significantly shorter running times.

\section{Monte Carlo algorithm}
\label{sec:mc}
In this section we describe the proposed Monte Carlo algorithm for sampling from the joint marginal posterior on $k$ and~$z$, Eq.~\eqref{eq:posterior}, with the uniform Dirichlet-categorical prior of Eq.~\eqref{eq:pzk}.  The algorithm also extends to the more general non-uniform prior of Eq.~\eqref{eq:nonempty} and we describe this extension in Appendix~\ref{app:genalpha}, but here we focus on the simpler uniform case for the sake of pedagogical clarity.

The algorithm is a form of collapsed Gibbs sampler similar in spirit to samplers for Dirichlet-process models such as those of \citet{Das14} and \citet{Khoufache23}, but applied to the Dirichlet-categorical prior and with some additional innovations that improve performance.  Given any initial choice of~$k$ and assignment~$z$ of the $N$ observations to components, the algorithm repeatedly performs the following actions.
\begin{enumerate}
\setlength{\itemsep}{4pt}
\item Choose a component~$r$ uniformly at random, then choose an observation~$i$ uniformly at random from that component.
\item Remove $i$ from component~$r$.
\item If $i$ was the only member of component~$r$, delete component~$r$, relabel component~$k$ to be the new component~$r$, and decrease~$k$ by~1.  (If $r=k$ then no relabeling is necessary; we simply delete the component.)
\item Assemble a set of $k+1$ candidate states, each of which has a weight associated with it, as follows.
\begin{enumerate}
\setlength{\itemsep}{4pt}
\item Of the $k+1$ states, $k$~of them are states in which $i$ is placed in one of the $k$ current components~$s$.  (We explicitly include the case $s=r$, where $i$ is placed back in the component~$r$ it was just removed from, in which case there is no overall change of state.)  The weights associated with each of these candidate states are
\begin{equation}
w_s = {P(x|k,z_{-i},z_i=s)\over P(x|k,z_{-i})},
\label{eq:a2w1}
\end{equation}
where $z_{-i}$ denotes the component assignments for all observations except for observation~$i$.
\item The last candidate state is one that makes $i$ the sole member of a new component~$k+1$ and increases $k$ by~1.  The weight associated with this state~is
\begin{equation}
w_{k+1} = {k^2\over N-k}\,
         {P(k+1) P(x|k+1,z_{-i},z_i=k+1)\over P(k) P(x|k,z_{-i})},
\label{eq:a2w2}
\end{equation}
where $P(k)$ is the prior on~$k$ as previously, and $k$ is the number of components before the new component is added.
\end{enumerate}
\item Once the set of target states is assembled, choose one of them with probability
\begin{equation}
p_s = {w_s\over\sum_s w_s}
\label{eq:pmu}
\end{equation}
and update the system to that new state.
\end{enumerate}
A proof of the correctness of this algorithm is given in Appendix~\ref{app:mcmc}.  A crucial feature of the algorithm is that we choose a \emph{component} in the first step, not an observation, then choose an observation from that component.  This process, if repeated alone without the other parts of the algorithm, results in assignments drawn exactly from the prior~$P(z|k)$, Eq.~\eqref{eq:pzk}, which is why no term for the prior appears in the formulas for the weights.  Moreover, these draws are ``rejection-free,'' meaning that no draws are discarded, as they are for instance in a Metropolis-Hastings process.  This change significantly improves the efficiency of the algorithm.  One way to understand the improvement is to notice that, if the prior were included in the weights, the factor of $n_r!$ in Eq.~\eqref{eq:pzk} would favor moving observations into large components, which would mean that an observation that started out in a large component would be more likely to be placed back in the same component again, resulting in a wasted move that did not update the state of the system.  Furthermore, the uniform selection of an observation at the start of each Monte Carlo move, as in a conventional Gibbs sampler, would result in a bias toward large components that would emphasize such wasted moves over moves that actually generate new states.  This issue has been noted by previous authors, such as \citet{MH18}, who suggest the split-merge sampling method of \citet{JN04} as a potential remedy.  Our algorithm, however, offers a simpler solution.  By moving the prior out of the weights and into the proposal mechanism, Monte Carlo moves that place observations in large components are no longer favored and the uniform selection of a random \emph{component} at the start of each move eliminates the selection bias in favor of large components.

As a specific example where this approach can make a difference, consider what happens when the algorithm creates a new component, which initially is small by definition.  Creation of components is a natural part of the operation of any algorithm that samples from the posterior distribution of the number of components, but inevitably the algorithm will sometimes create components where they are not justified by the data, in which case they will normally be immediately deleted again.  With a traditional Gibbs sampler (or collapsed Gibbs sampler), the process of deletion is slow because an algorithm that updates all observations with equal frequency updates the observations in a small component of size~$\Ord(1)$ on only a fraction $\Ord(1/N)$ of Monte Carlo steps.  Our algorithm, on the other hand, since it chooses a random component on every step, performs such updates $\Ord(1/k)$ of the time, and hence can be expected to remove unwanted components a factor of $\Ord(N/k)$ faster than the traditional algorithm.  For large data sets this factor can be on the order of thousands or more.

It would be possible to speed up our algorithm further by parallelization.  The bulk of the computational effort goes into calculating the weights in Eqs.~\eqref{eq:a2w1} and \eqref{eq:a2w2}, which has to be done anew for every Monte Carlo step.  The individual weights however are not dependent and can be calculated in any order, which makes the algorithm trivially parallelizable and an ideal candidate for a multithreaded or vectorized implementation.

\subsection{Parameter values}
Because the parameters~$\theta$ are integrated out of the marginal likelihood in Eq.~\eqref{eq:marginal}, we do not get estimates of them directly from the Monte Carlo.  However, it is straightforward to estimate them for a given component assignment~$z$ and number of components~$k$ by writing
\begin{equation}
P(\theta|x,k,z) = {P(x|k,z,\theta) P(\theta|k)\over P(x)},
\label{eq:esttheta}
\end{equation}
with $P(x|k,z,\theta)$ as in Eq.~\eqref{eq:likelihood}.  If $P(\theta|k)$ is the appropriate canonical prior, Eq.~\eqref{eq:esttheta} takes the same functional form as the prior, which generally makes it straightforward to estimate the expectation of~$\theta$, its variance, modal value, and so forth.  These estimates do still depend on~$k,z$ and our Monte Carlo procedure returns many candidate values of~$k,z$ and hence many values of~$\theta$.  For practical purposes one would often prefer just a single ``best'' value, which leads us to our next topic.

\subsection{Label switching and consensus components}
An issue with sampling from mixture models is that the component labels themselves are meaningless.  If we take a component assignment~$z$ and permute the $k$ component labels we still have the same division of observations, and hence all permutations have the same posterior probability.  This means for instance that the marginal probability $P(z_i=r|x,k)$ of observation~$i$ belonging to component~$r$ is always a constant $1/k$ independent of the data~$x$, so one cannot meaningfully ask ``What component does $i$ belong to?''  The components are not identifiable.  This problem is manifested in the Monte Carlo results as ``component switching'' or ``label switching''---the algorithm may sample the same or similar component assignments but with different permutations of the labels, so that the similarity is difficult to see.

The method has not failed.  It is drawing correctly from the posterior distribution, but for practical purposes it is not giving us the type of answer we are looking for.  In most cases the practitioner would prefer a single unambiguous assignment of observations to components, or at least a set of closely similar ones.  Methods that employ EM algorithms deal with this by selecting a single estimate of the parameters~$\theta$ and then reporting component assignments conditioned on those values.  This breaks the symmetry over permutations, but the EM algorithm has other shortcomings, as discussed in the introduction.

What can we do instead?  Some investigators have suggested imposing symmetry-breaking rules on the component assignments so that only one assignment in each orbit of the permutation group is sampled \citep{RG97,Stephens00a}, or choosing the permutations that make the sampled assignments most similar \citep{ZDA23}.  These approaches work well when the posterior distribution is concentrated around the modal component assignment, but fail when the distribution is broad enough that its faithful representation would require frequent label switching.  In the latter situation, algorithms that prohibit label switching give biased estimates \citep{Gelman13}.

A better approach, in our opinion, is to sample assignments without restriction, allowing label switching, and then, if a single assignment is desired, to construct one after the fact as the \textit{consensus clustering} that best captures the common features of all the samples.  Many methods for doing this have been proposed \citep{MPMG03,Bryant03,GF08,VR11,LF12,ZDA23} and we give an example application in Section~\ref{sec:lca}.  It is worth noting, however, that it may not be possible to capture all the salient features of the sampled assignments in a single consensus configuration---there may be multiple configurations that have high posterior probability but which differ significantly.  This kind of behavior has been observed for instance in network community detection \citep{GDC10}, and has led to the development of alternative approaches, such as ``building block'' decompositions \citep{RN20,Arthur24} or methods for finding multiple representative assignments within a large sample \citep{KN22}.

Alternatively, we can simply restrict ourselves to analyses based purely on quantities that are invariant under permutations of the component labels.  Examples of such quantities include the coincidence rate---the frequency with which two observations belong to the same component---and the mutual information between sampled component assignments for two observations, or between the component assignments and the data.  There are problems of interest that can be tackled directly using such quantities.  An example is variable selection, the problem of deciding which observed variables are most informative about component membership.  We give an example in Section~\ref{sec:lca}.

\section{Results}
In this section we give results on applications and performance of our methods for a variety of models.

\subsection{Gaussian mixtures}
\label{sec:gaussian}
For our first example, we perform benchmark tests using perhaps the simplest of possible models, a univariate Gaussian mixture with constant variance.  This model is fully parametrized by the means $\mu_1\ldots \mu_k$ of the components, plus the number of components~$k$ and assignments~$z$.  The data likelihood is
\begin{align}
P(x|k,z,\mu) 
  &= \prod_i {1\over\sqrt{2\pi\sigma^2}}
     \exp\biggl[ -{(x_i-\mu_{z_i})^2\over2\sigma^2} \biggr] \nonumber\\
  &= (2\pi\sigma^2)^{-N/2} \prod_r
     \exp\biggl[ -{\sum_{i\in r} (x_i-\mu_r)^2\over2\sigma^2} \biggr] \nonumber\\
  &= (2\pi\sigma^2)^{-N/2} \prod_r
     \exp\biggl[ -{n_r\over2\sigma^2} \bigl[(\av{x}_r-\mu_r)^2 + \sigma_r^2\bigr]
     \biggr],
\end{align}
where $\av{x}_r$ and $\sigma_r^2$ denote the mean and variance of the observations within component~$r$, and $n_r$ is the number of observations in component~$r$ as previously.  Integrating with respect to~$\mu_r$ over an interval of width~$a$, with $a\gg2\pi\sigma^2$ and a uniform prior, the marginal likelihood is then
\begin{equation}
P(x|k,z) = {1\over a^k(2\pi\sigma^2)^{(N-k)/2}} \prod_r {1\over\sqrt{n_r}}\,
   \e^{-n_r\sigma_r^2/2\sigma^2}.
\end{equation}
For the prior~$P(k)$ on the number of classes we make the uniform choice~$P(k)=1/N$, then the weight of Eq.~\eqref{eq:a2w1} for Monte Carlo moves that do not create a new component becomes
\begin{equation}
w_s = \sqrt{n_s/n_s'\over 2\pi\sigma^2} \,
      \e^{-(n_s'{\sigma_s'}^2-n_s\sigma_s^2)/2\sigma^2}.
\end{equation}
With a little extra work we can show that
\begin{equation}
n_s'{\sigma_s'}^2-n_s\sigma_s^2 = {n_s\over n_s'} (x_i - \av{x}_s)^2,
\end{equation}
and putting $n_s' = n_s+1$, we have
\begin{equation}
w_s = \sqrt{n_s/(n_s+1)\over 2\pi\sigma^2} \,
        \exp \biggl[ -{n_s/(n_s+1)\over2\sigma^2} (x_i-\av{x}_s)^2 \biggr].
\label{eq:gaussw1}
\end{equation}
Performing the same series of steps for Eq.~\eqref{eq:a2w2}, the weight for the remaining move which creates a new component $k+1$ is simply
\begin{equation}
w_{k+1} = {k^2/a\over N-k}.
\end{equation}

This simple model provides a convenient arena for benchmarking the performance of our algorithm against competing methods.  For these tests we generate synthetic data sets of $N=10\,000$ observations each, divided equally between $k$ Gaussian components with the observations in component $r=1\ldots k$ drawn from $x_i \sim \text{Normal}(3k,1)$.  The performance of Monte Carlo algorithms for this problem typically varies with~$k$, either because of the time taken for individual Monte Carlo steps or because of varying mixing times, or both, so we report results for a range of $k$ values to give a complete picture.

We compare the performance of our algorithm against three other popular Monte Carlo methods for Bayesian mixture models, as implemented in the software package BayesMix \citep{BGGG25}: Neal's second and third algorithms, which are traditional Gibbs samplers \citep{Neal00} and the split-merge algorithm of \citet{JN04}.  Results are given in Table~\ref{tab:benchmarks}.

\begin{table}
\caption{Running time per sweep in milliseconds and estimated mixing time in sweeps for the algorithm of this paper and a selection of competing algorithms, applied to a Gaussian model with synthetic data and various numbers of components~$k$.  Figures in parentheses indicate standard errors on the trailing digits.}
\label{tab:benchmarks}
\setlength{\tabcolsep}{4pt}
\begin{tabular}{r|cc|cc|cc|cc}
    & \multicolumn{2}{c|}{This paper} & \multicolumn{2}{c|}{Neal's algorithm 2} & \multicolumn{2}{c|}{Neal's algorithm 3} & \multicolumn{2}{c}{Split-merge} \\
$k$ & Time      & Mixing            & Time      & Mixing          & Time      & Mixing         & Time      & Mixing \\
\hline
3   & 1.4        & 21.2(8.4)         & 17.9       & 120.4(11.7)     & 23.6       & 144.9(16.7)    & 139.3      & 18.7(2.5) \\
5   & 2.2        & 25.4(6.2)         & 27.0       & 79.8(10.3)      & 23.8       & 84.2(11.7)     & 131.2      & 23.9(2.2) \\
7   & 3.7        & 20.8(3.3)         & 29.2       & 80.9(7.9)       & 23.9       & 60.4(3.5)      & 129.6      & 11.2(1.3) \\
10  & 4.5        & 24.0(2.7)         & 34.1       & 49.9(3.4)       & 22.5       & 43.5(5.1)      & 152.8      & 8.1(5)    \\
\end{tabular}
\end{table}

BayesMix is written in the C\texttt{++} language for speed and is a well optimized piece of software, but it can only be as efficient as the algorithms it implements, and in this respect it is laboring under two disadvantages.  First, the algorithms are complex.  Even a single Monte Carlo step can require a large amount of computation.  For instance, the fastest of the competing algorithms we consider is Neal's second algorithm.  In our tests we measure progress in ``sweeps,'' meaning $N$ Monte Carlo updates of individual observations, so that each observation is updated once on average per sweep.  For the case where $k=5$, a complete sweep of Neal's second algorithm takes 27.0 milliseconds of CPU time on the hardware used for these tests.  For the algorithm of this paper, by contrast, a complete sweep takes 2.2 milliseconds on the same hardware, less than a tenth of the time.  Not all of this difference will be due to the complexity of the algorithm.  Some is due to complexity of the models themselves or to overheads inherent in the architecture of the BayesMix package, although the authors state that such overheads are small.  Nonetheless, for a practitioner applying these methods the differences are real and will have a substantial impact on running times.

The second disadvantage of these algorithms is that their mixing time can be slow, meaning it takes many successive Monte Carlo sweeps to generate one statistically independent sample of the system.  The number of sweeps needed to generate an independent sample can be quantified by calculating the correlation time of the sample chain.  In this paper we compute the autocorrelation of the log-likelihood and then from it calculate the integrated correlation time \citep{NB99}.  Figures for the resulting times, measured in sweeps and averaged over ten repetitions with randomly generated data, are given in Table~\ref{tab:benchmarks}.  Here again we see significant differences between algorithms.  Once more taking the example of Neal's second algorithm for the case $k=5$, the correlation time is measured to be about 80 Monte Carlo sweeps, while for the algorithm of this paper it is about 25 sweeps, less than a third as much, meaning that it takes significantly fewer steps to generate one independent sample configuration.  The combination of faster sweeps and faster mixing makes the algorithm of this paper about 38 times faster overall than Neal's algorithm in this particular test, enough to make the difference between a calculation that runs in an hour and one that takes just a minute or two.  A difference of this magnitude is not merely a quantitative one, but a qualitative one also: the ability to obtain useful results in minutes or seconds makes possible forms of interactive and exploratory data analysis that are impractical with slower methods.

For the other algorithms the details vary but the overall outcome is similar.  Consider, for instance, the split-merge algorithm.  This powerful algorithm uses nonlocal moves that split and merge whole components to achieve in a single Monte Carlo step what in other algorithms takes many steps.  This improves mixing time, but at the expense of greater computational effort.  Looking again at the case of $k=5$, the correlation time for the split-merge algorithm is significantly better than Neal's algorithm---only 24 sweeps, which is also slightly better than our own algorithm at 25 sweeps.  On the other hand, each sweep of the system with $N=10\,000$ observations takes 131 milliseconds for the split-merge algorithm, compared with just 2.2 milliseconds for our algorithm, so that our algorithm is still faster overall by a factor of about~56.

\subsection{Poisson mixtures}
\label{sec:poisson}
For our next example we consider a mixture of Poisson distributions over integer observations~$x_i$, such that the data likelihood is
\begin{equation}
P(x|k,z,\mu) = \prod_i {\mu_{z_i}^{x_i}\over x_i!} \exp(-\mu_{z_i}) \nonumber\\
  = \biggl[ \prod_i {1\over x_i!} \biggr]
    \biggl[ \prod_{r=1}^k \mu_r^{\sum_{i\in r} x_i} \e^{-n_r\mu_r} \biggr].
\end{equation}
Again the model is completely parametrized by the means~$\mu_r$ of the components, along with the number~$k$ of components and the assignments~$z_i$.  Assuming a conventional gamma prior on the means of the form
\begin{equation}
P(\mu) = {\beta^\alpha\over\Gamma(\alpha)} \mu^{\alpha-1} \e^{-\beta\mu}
\end{equation}
and integrating, we find the marginal likelihood to be
\begin{equation}
P(x|k,z) = \prod_i {1\over x_i!} \prod_{r=1}^k {\beta^\alpha \Gamma(\sum_{i\in r} x_i+\alpha)\over\Gamma(\alpha) (n_r+\beta)^{\sum_{i\in r} x_i + \alpha}}.
\end{equation}
Substituting this expression into Eqs.~\eqref{eq:a2w1} and~\eqref{eq:a2w2}, we get Monte Carlo weights
\begin{equation}
w_s = {\Gamma(X_s+x_i+\alpha)\over\Gamma(X_s+\alpha)}\,
      {(n_s+\beta)^{X_s+\alpha}\over(n_s+\beta+1)^{X_s+x_i+\alpha}},
\qquad
X_s = \sum_{i\in s} x_i
\end{equation}
for steps that move an observation~$i$ to an existing component~$s$, and
\begin{equation}
w_{k+1} = {k^2\over N-k}\, {P(k+1)\over P(k)}\,
         {\Gamma(x_i+\alpha)\over\Gamma(\alpha)}\,
         {\beta^\alpha\over(\beta+1)^{x_i+\alpha}}
\end{equation}
for steps that move observation~$i$ to newly created component~$k+1$.

\begin{figure}
\begin{center}
\includegraphics[width=9cm]{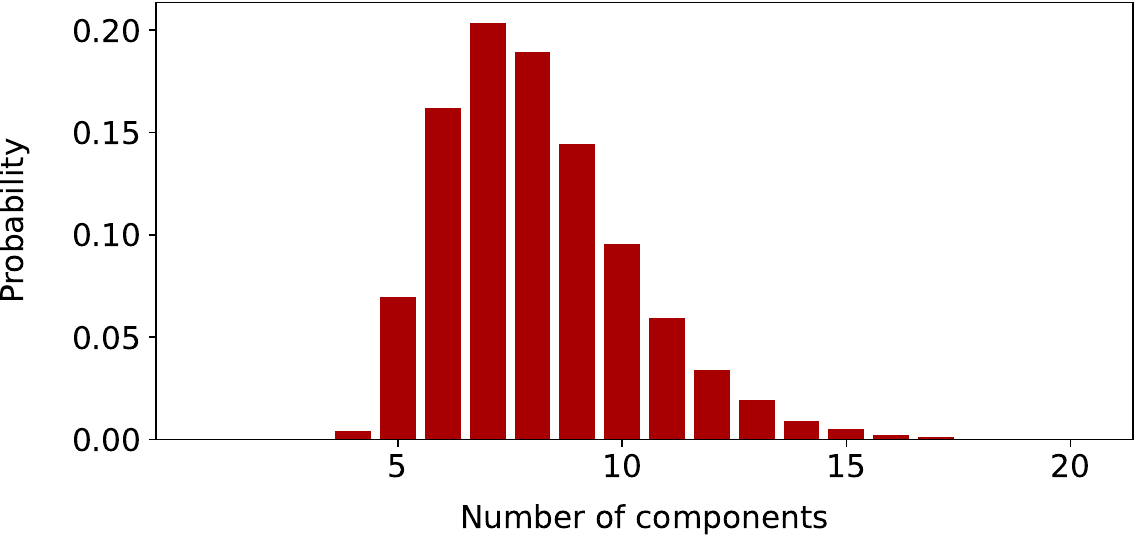}
\end{center}
\caption{Estimated distribution of the number of components in a fit of the epileptic seizures data set of \citet{WPCL96} to the Poisson mixture model described in the text.}
\label{fig:seizures}
\end{figure}

As an example, Fig.~\ref{fig:seizures} shows results for a clinical data set from \citet{WPCL96} recording the number of seizures experienced per day by an epileptic patient under varying treatment regimes over a 140-day observation period.  Wang~\textit{et al.}~analyzed the data using two- and three-component Poisson mixture models.  For our analysis we applied the Monte Carlo algorithm above with $\alpha=1$ and $\beta=0.01$ for 10\,000 sweeps of burn-in followed by 100\,000 sweeps for sampling, the entire calculation taking about three seconds on the author's laptop.  Figure~\ref{fig:seizures} shows the estimated posterior distribution over the number of components, which is simply a histogram of the sampled values of~$k$ and, as the figure reveals, fits with both two and three components are firmly ruled out: out of 100\,000 samples, zero of them had either $k=2$ or $k=3$.  The smallest observed number of components was $k=4$ and the smallest plausible value (at a $p=0.05$ significance level) was $k=5$.  The modal value was $k=7$.  Realistically, this implies one of two things: either the original authors' assumption of two or three components was wrong or the daily seizure events are not Poisson distributed and a mixture of Poissons is a poor model.

\begin{figure}
\begin{center}
\includegraphics[width=11cm]{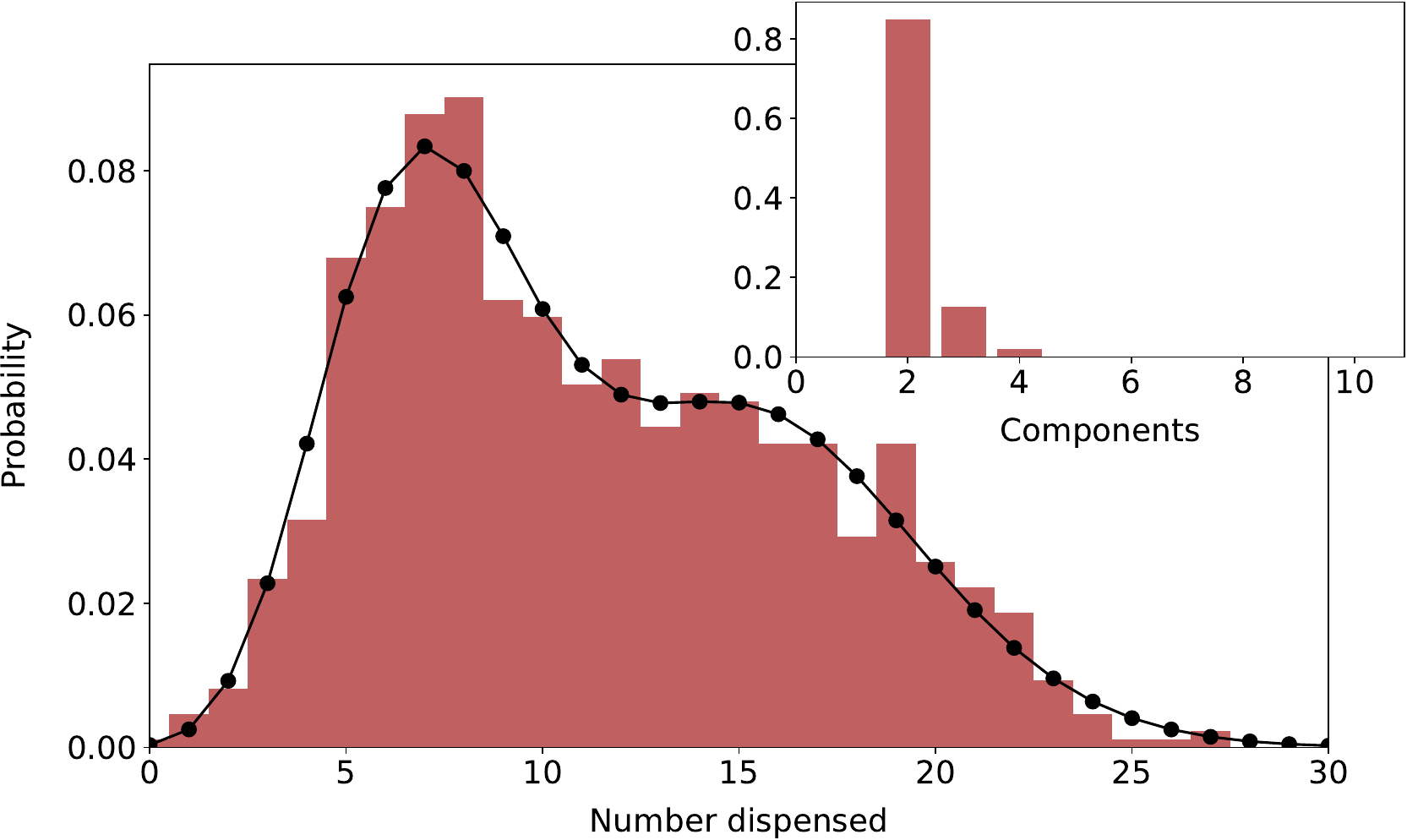}
\end{center}
\caption{Main figure: The empirical distribution (histogram) and fitted distribution (points) for the number of candies dispensed by the machine.  Inset: The posterior distribution of the number of components estimated from the Monte Carlo results.}
\label{fig:candies}
\end{figure}

Figure~\ref{fig:candies} shows another, more whimsical application, to a data set that comes from a candy dispenser in the hallway outside the author's office---a machine that dispenses candy at the push of a button.  Allegations have swirled for some time that this machine is not fair: sometimes (it is alleged) it dispenses a generous handful of candies, sometimes a miserly pittance.  The data set represents real counts of the number of candies dispensed on 853 pushes of the button, which were analyzed using the Poisson model above with 10\,000 sweeps of burn-in followed by 100\,000 sweeps of sampling, which takes about six seconds of running time.  The figure shows the distribution of numbers of candies (main figure) and the inferred number of components (inset).  As the inset shows, the results firmly exclude the possibility that there is only one component and lean heavily into the hypothesis that there are two: a miserly one (mean 7.382(2), 54.7\% of observations) and a generous one (mean 15.987(3), 45.3\% of observations).  The main figure shows the observed distribution of numbers of candies along with the fitted distribution inferred from the model.  Together these results strongly support the charge that the machine is inequitable in the performance of its duties.

\subsection{Latent class analysis}
\label{sec:lca}
For a more substantial test of our algorithm, we turn to latent class analysis (LCA), the mixture modeling of categorical data, such as is used in the analysis of test results or survey data \citep{Goodman74,McCutcheon87,BR93,LLSL18}.  Here we discuss LCA in the language of survey data, but the methods we describe are broadly applicable to any categorical data.

Consider a survey in which $N$ respondents are divided into $k$ components, also called \textit{classes} in this context, and give responses to $Q$ multiple-choice questions.  Let $\theta_{rqx}$ be the probability that a respondent in class~$r$ answers question~$q$ with response~$x$.  If $x_{iq}$ is the answer given by respondent~$i$ to question~$q$, and assuming independent responses, the complete data likelihood is
\begin{equation}
P(x|k,z,\theta) = \prod_{iq} \theta_{z_iqx_{iq}}
  = \prod_{rqx} \theta_{rqx}^{m_{rqx}},
\label{eq:lcalike}
\end{equation}
where $m_{rqx}$ is the number of respondents in class~$r$ who answer question~$q$ with response~$x$.  One normally assumes a symmetric Dirichlet prior on the $\theta_{rqx}$ for each class/question combination, with some concentration parameter~$\eta$, which gives
\begin{equation}
P(x,\theta|k,z,\eta)
  = \prod_{rq} {\prod_{x=1}^{k_q} \theta_{rqx}^{m_{rqx}+\eta-1}\over
    \mathrm{B}(\eta,\eta,\ldots)},
\label{eq:paomega}
\end{equation}
where $k_q$ is the number of distinct possible answers to question~$q$ and $\mathrm{B}(\eta,\eta,\ldots)$ is the multivariate beta function again.  Integrating over the $\theta$ parameters, we then have
\begin{align}
P(x|k,z,\eta) &= \int P(x,\theta|k,z,\eta) \>d\theta
   =\prod_{rq} {1\over\mathrm{B}(\eta,\eta,\ldots)} \int
     \prod_{x=1}^{k_q} \theta_{rqx}^{m_{rqx}+\eta-1} \>d\theta \nonumber\\
  &= \prod_{rq} {\Gamma(\eta k_q)\over\Gamma(n_r+\eta k_q)}
     \prod_{x=1}^{k_q} {\Gamma(m_{rqx}+\eta)\over\Gamma(\eta)},
\label{eq:pakz}
\end{align}
where we have made use of the fact that $\sum_x m_{rqx} = n_r$ for all~$q$.  Combining this expression with Eqs.~\eqref{eq:a2w1} and~\eqref{eq:a2w2}, the weights for our Monte Carlo algorithm for LCA are
\begin{equation}
w_s = \prod_q {m_{sqx_{iq}}+\eta\over n_s+\eta k_q}, \qquad
w_{k+1} = {k^2\over N-k} {P(k+1)\over P(k)} \prod_q {1\over k_q},
\end{equation}
where all quantities, including~$k$, denote values before observation~$i$ is placed in its new class.

\subsubsection{Synthetic tests}
\label{sec:synthetic}
As a first test of this algorithm we benchmark it against synthetic data generated from the same model.  These calculations can be thought of as consistency tests and we focus particularly on whether we are able to correctly recover the number of classes present in the data.

\begin{figure}
\begin{center}
\includegraphics[width=6.7cm]{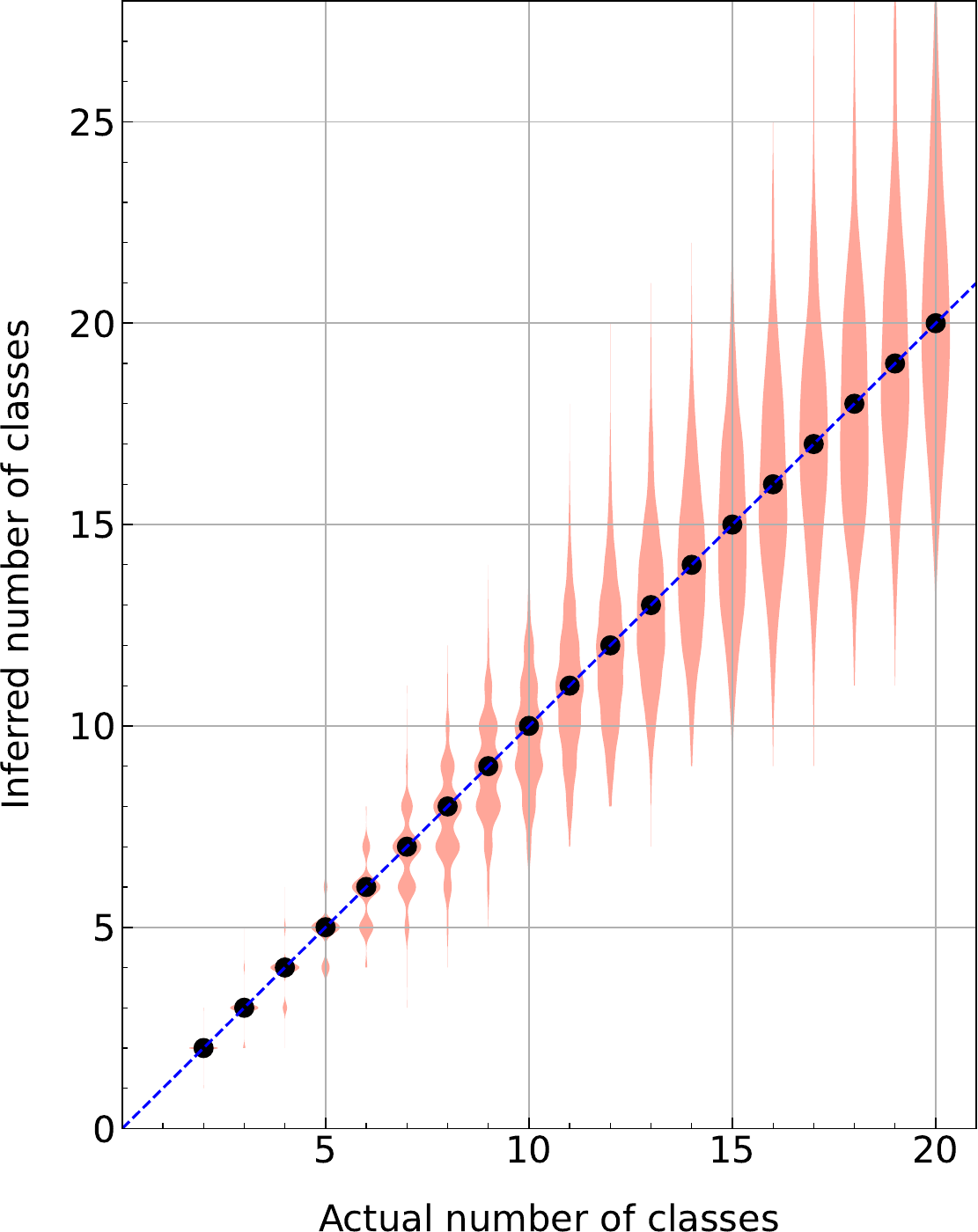}
\end{center}
\caption{Inference of the number of classes in synthetic data.  For each value of the number of classes~$k$ from 2 to~20, we generate 1000 synthetic data sets as described in the text and attempt to infer the number of classes using the method of this paper.  The violin plots indicate the range of the results and the points indicate the median inferred number of classes.  When inferred and actual numbers agree, the points lie on the diagonal line.}
\label{fig:violin}
\end{figure}

The synthetic data sets we generate have $N=1000$ respondents, $Q=10$ questions, $k_q=4$ possible answers to each question, and a Dirichlet concentration parameter of $\eta=1$.  Figure~\ref{fig:violin} summarizes results from application of our Monte Carlo algorithm to a large number of such data sets with varying numbers of classes~$k$ from 2 up to~20.  For each value of $k$ we generate 1000 data sets and use our Monte Carlo algorithm to compute the MAP estimate of~$k$ for each one, which is simply the most commonly occurring value of $k$ among all samples drawn.  Figure~\ref{fig:violin} shows the results plotted against the ground-truth values used to generate the data.  In this figure the violin plots represent the variation in $k$ across all 1000 repetitions and the circles represent the median values.  If the algorithm has learned the number of classes correctly the results should fall on the diagonal dashed line and, as the figure shows, the median results are perfect in this case.  There is, however, considerable variation about the median.  This is typical of results from these methods and should be considered a strong point of the approach.  Even for synthetic data there is genuine uncertainty in the data about the number of classes, which the Bayesian approach reveals in a way that other methods cannot.

\subsubsection{CBS/New York Times opinion poll}
Turning to real-world data sets, we give three examples of applications drawn from different spheres.  The data sets are described in detail in Appendix~\ref{app:data}.

Our first example is a classic survey data set from a CBS/New York Times opinion poll of 1566 respondents in the United States, taken in September 2011, during the first term of the administration of president Barack Obama, and asking about a range of then-topical issues, including the president's job approval, the economy, government spending, taxes, and health care, as well as basic demographics such as respondents' sex and marital status.

\begin{figure}
\begin{center}
\includegraphics[width=\textwidth]{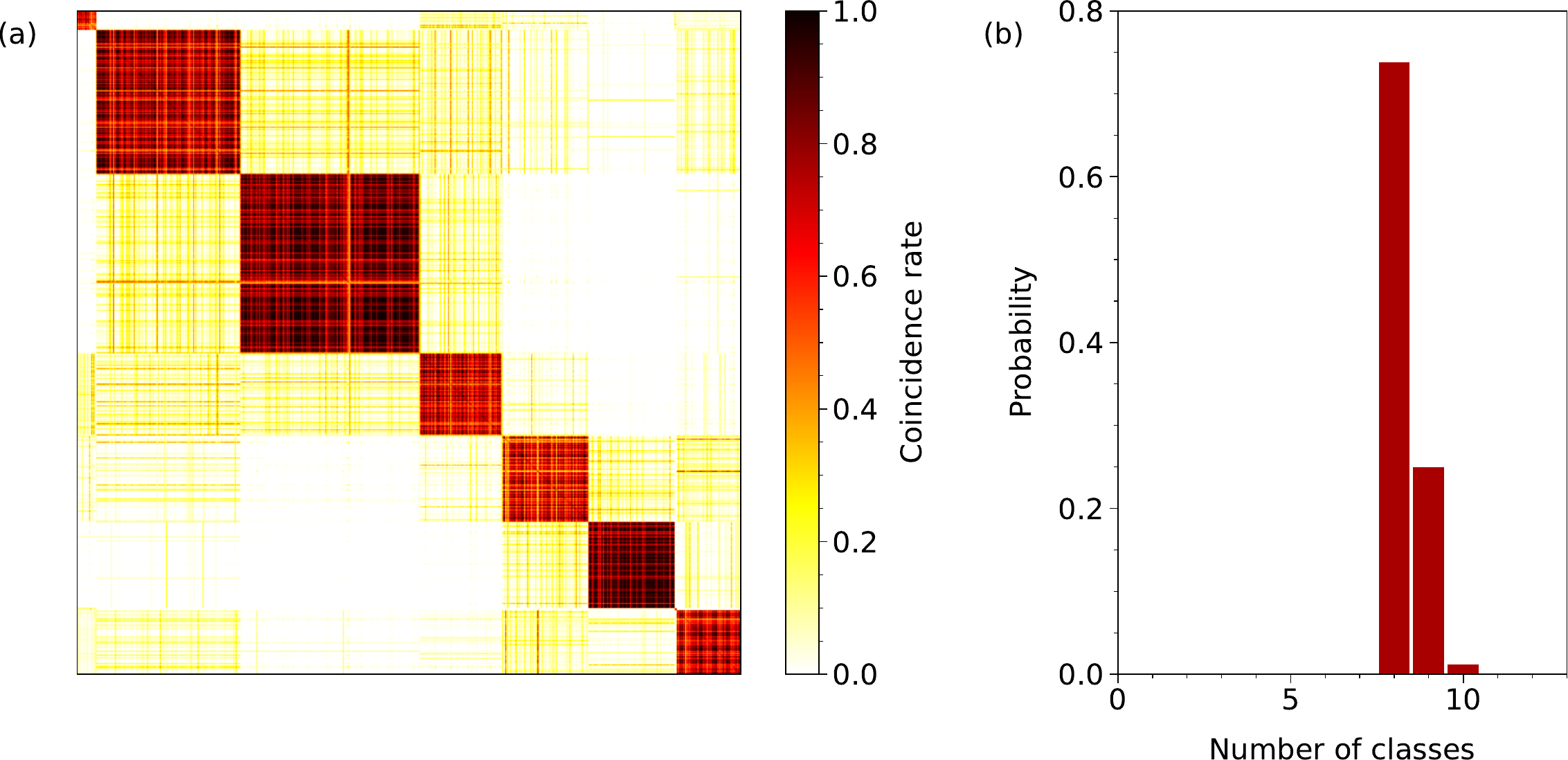}
\end{center}
\caption{(a)~Consensus matrix of class assignments for the 1566 respondents in the CBS/New York Times poll.  The rows and columns of the matrix have been permuted to place the members of each class in a contiguous block and make the class structure clearer to the eye.  (We perform a similar operation for the other data sets in this section also.)  (b)~The distribution of sampled values of the number of classes~$k$, which provides an estimate of the posterior probability distribution over~$k$.}
\label{fig:cbs}
\end{figure}

We perform a single Monte Carlo run on this data set of 2500 sweeps for burn-in followed by 25\,000 sweeps to take samples, which produces the results shown in Fig.~\ref{fig:cbs}.  Panel~(a) shows the \textit{consensus matrix} (also sometimes called the association matrix), the matrix whose elements are the coincidence rates (the fraction of samples in which a pair of respondents are in the same class together).  In this case the algorithm has identified eight distinct classes in the data, as indicated by the prominent blocks along the diagonal of the consensus matrix, although one class, represented by the second-to-last block in the matrix, is small, with only five members, making it more difficult to see.  Figure~\ref{fig:cbs}b shows the posterior distribution over the number of classes and we see that the method is confident in this case in its conclusion that there are eight classes in the data (probability~0.74).

\subsubsection{Diagnostic survey for Alzheimer's disease}
\begin{figure}
\begin{center}
\includegraphics[width=\textwidth]{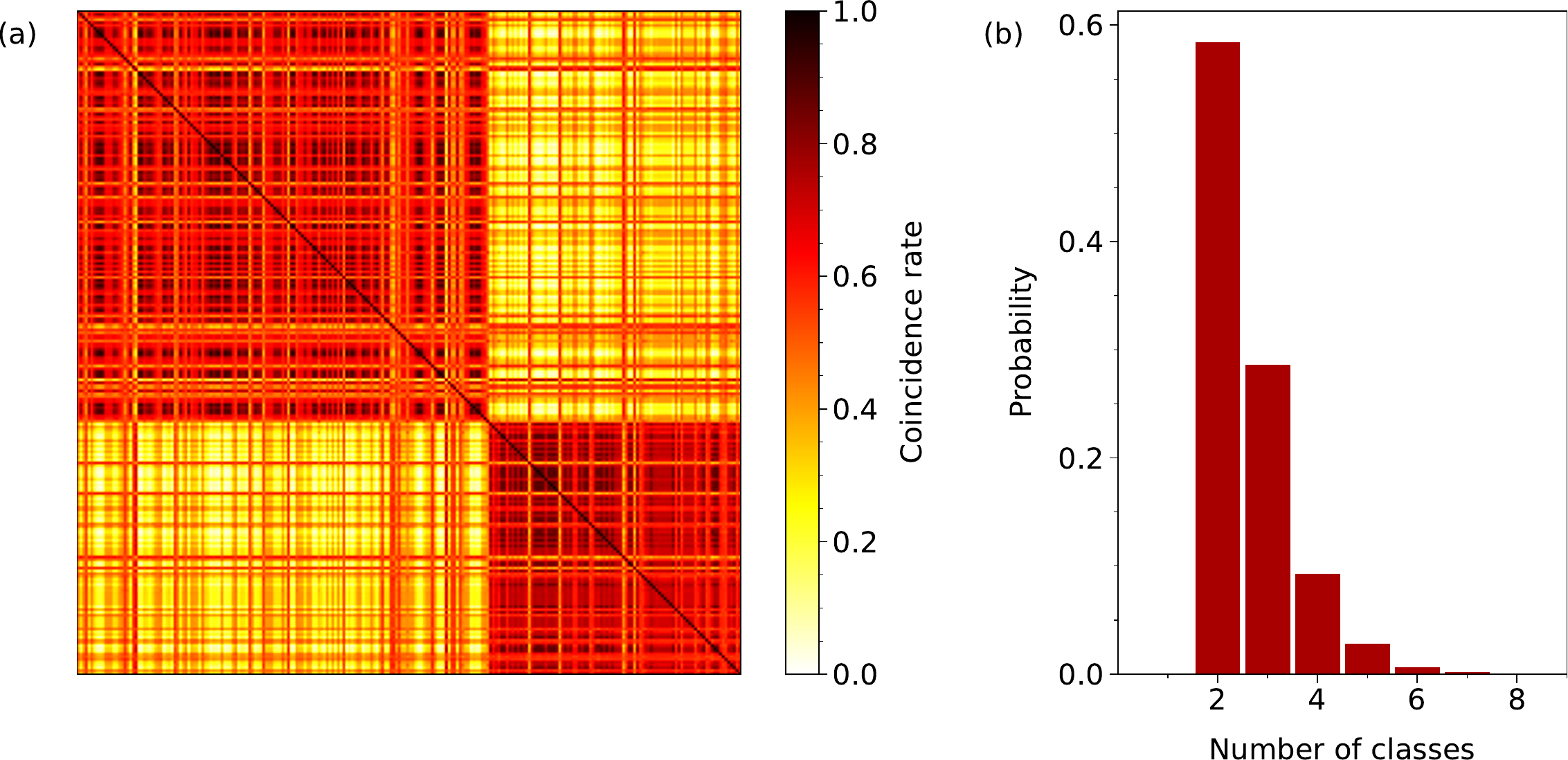}
\end{center}
\caption{(a)~Consensus matrix for the patients in the cognitive function data set, which shows two classes.  (b)~The distribution of number of classes in the Monte Carlo sample.  Although $k=2$ is the most likely (MAP) value, there is some uncertainty in this case.}
\label{fig:alzheimers}
\end{figure}

Our next example is a clinical one: we analyze data from Moran and Walsh (\citeyear{Moran04}; \citealt{Walsh06}) describing observations of potential cognitive decline in 240 patients at a memory clinic in Ireland.  The survey instrument asked primary caregivers about the presence or absence in the patients of six symptoms commonly associated with Alzheimer's disease: activity disturbance, affective disorder, aggression, agitation, diurnal rhythm disturbance, and hallucinations.  Figure~\ref{fig:alzheimers} shows the consensus matrix and distribution of the number of classes for these data and, as we can see, there are most likely two groups in the data, although the division between the classes is less clear than for the previous example and the posterior distribution on~$k$ allows for values up to $k=7$.  \citet{Walsh06} also found two classes in his work, corresponding to patients with and without substantial symptoms, although he comments that there is potentially interesting additional structure in the three-class division, which carries a nontrivial posterior probability in our analysis (Fig.~\ref{fig:alzheimers}b).  \citet{WWM16}, analyzing the same data using a Monte Carlo method based on the allocation sampler of \citet{NF07}, reached similar conclusions, although their calculation took over 3 minutes, where ours takes a fraction of a second.

\begin{figure}
\begin{center}
\includegraphics[width=9cm]{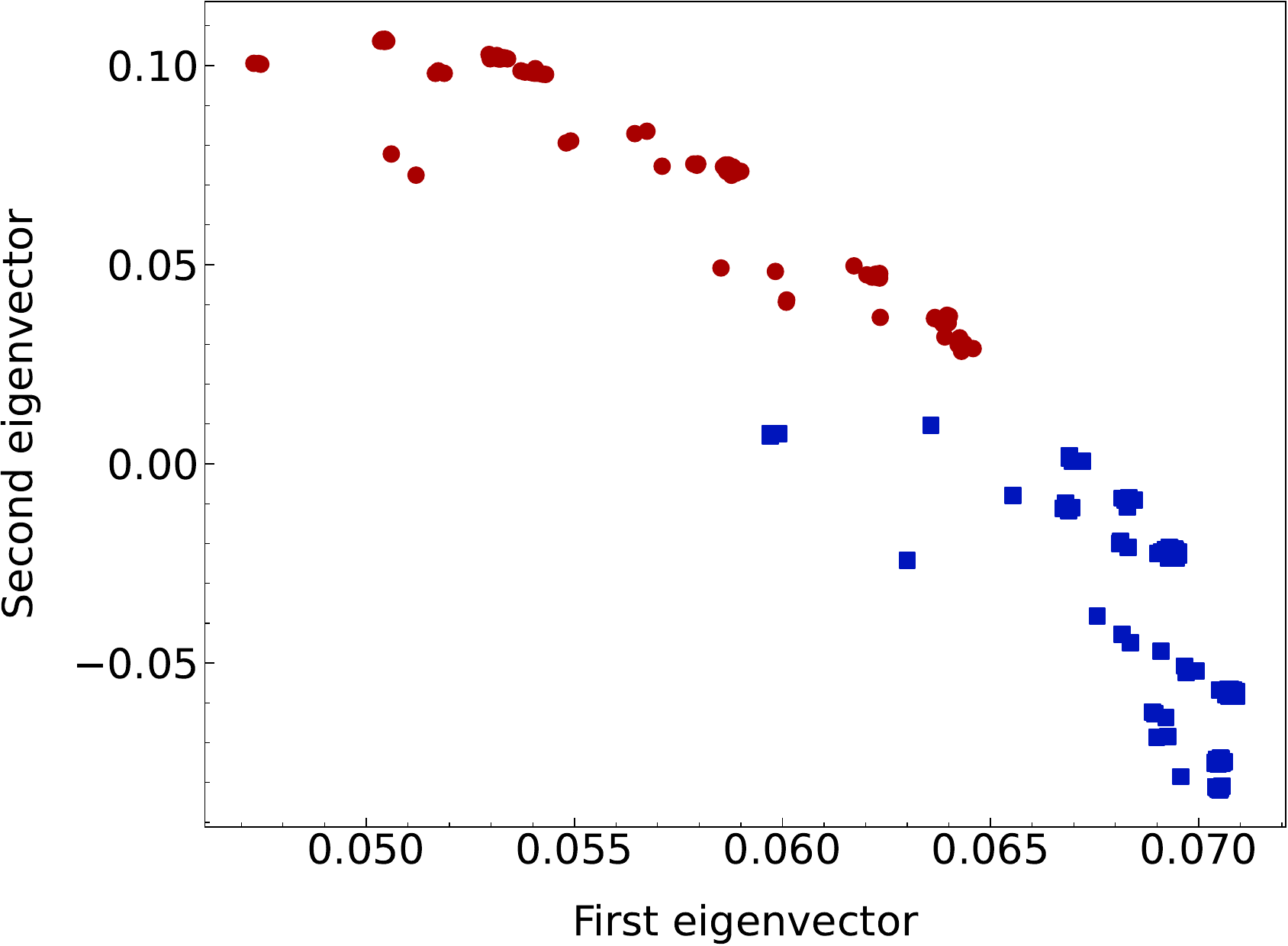}
\end{center}
\caption{The patients in the cognitive test data set, arranged according to the leading two eigenvectors of the consensus matrix.  The shapes of the points represent the results of a simple partition into two classes using $k$-means clustering.}
\label{fig:scatter}
\end{figure}

The roughly block-diagonal form of the consensus matrix seen here and in our other examples implies that the matrix is approximately of low rank, which suggests a possible further analysis: we could look at the leading eigenvectors of the matrix in a manner akin to principal component analysis or correspondence analysis.  We do this in Fig.~\ref{fig:scatter} and the figure does a good job of revealing structure and clustering in the data, which in turn suggests a natural method of consensus clustering in which we group the points on such a plot using, for example, $k$-means clustering, with the number of classes chosen to match the MAP estimate from the Monte Carlo results (which is two in this case).  This approach has been studied by \citet{ZDA23}, who show that it has provably good performance under a simple model of label perturbation.

Applying this method in the present case gives the division shown by the shapes/colors of the points in the figure and produces a sensible consensus.  On the other hand, it also reveals why the Monte Carlo was less certain of the number of classes in this case: we can clearly see that there are subclusters within each of the larger groups that could, within reason, be considered to be classes in their own right.

\subsubsection{Geographic and socio-political traits of the 50 United States}
\label{sec:states}
Our last example is an application to non-survey data.  This data set is a geographic one that focuses on the 50 United States and records 12 characteristics for each one, including their position on a range of hot-button issues such as abortion access, gun laws, and the death penalty, as well as more neutral variables such as weather, educational attainment, and whether the state has a professional American football team.  The full list of variables is given in Table~\ref{tab:states}.

\begin{figure}
\begin{center}
\includegraphics[width=\textwidth]{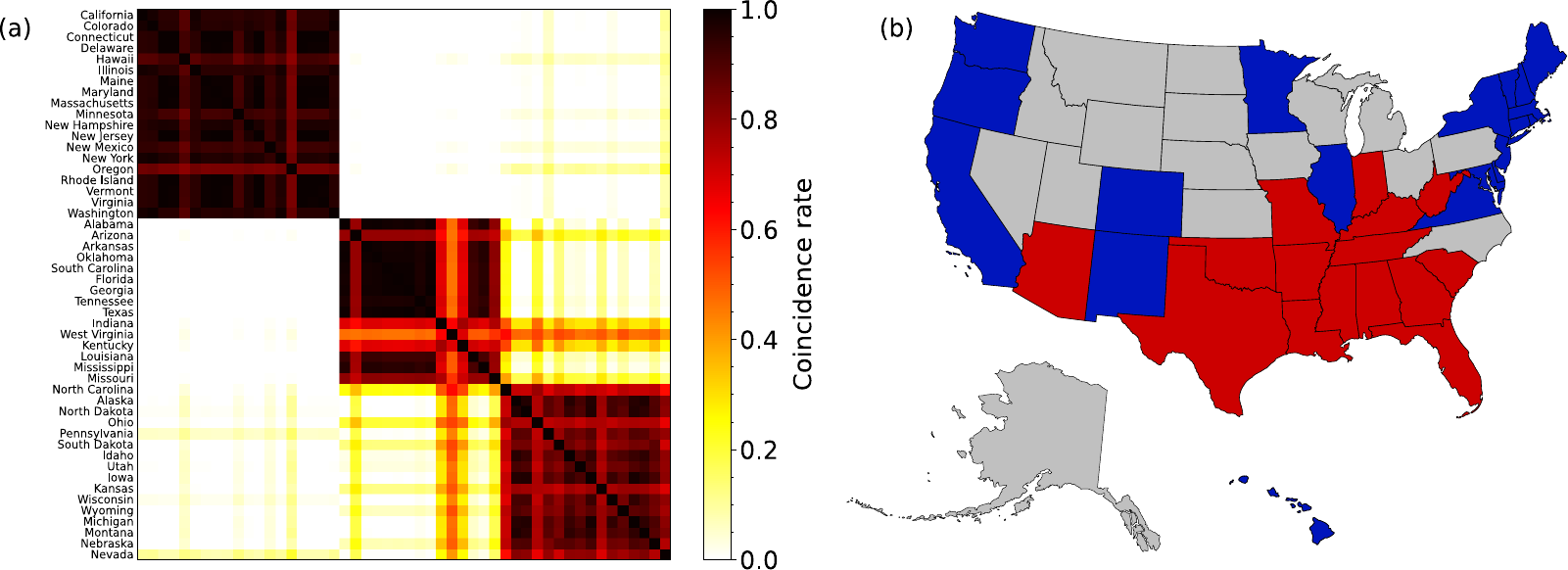}
\end{center}
\caption{(a)~Consensus matrix for the classification of the 50 United States, which finds three clear classes, as indicated on the map~(b).  One class, shown in blue, corresponds to traditionally left-leaning states and another, in red, to right-leaning ones.  The class shown in gray consists of states that implement a mix of left- and right-leaning policies, such as Ohio, whose constitution simultaneously guarantees a right to abortion and forbids same-sex marriage.}
\label{fig:states}
\end{figure}

On this comparatively small data set our method runs quickly, completing 25\,000 Monte Carlo sweeps in under a second and finding three clear classes, as shown in Fig.~\ref{fig:states}a.  The classes correspond to recognized political divisions among the states: there is a group of left-leaning states (so-called ``blue states''), a group of right-leaning ones (``red states''), and a group whose positions on the issues are a mix of left and right.  Figure~\ref{fig:states}b shows the division on a map.

\begin{table}
\caption{Summary of the 12 variables used in the 50-states data set.  Further details are given in Appendix~\ref{app:data}, Section~\ref{app:states}.}
\label{tab:states}
\begin{tabular}{ll}
Variable & Definition \\
\hline
Abortion      & Legality of abortion \\
Cannabis      & Legality of cannabis \\
Census region & Geographic location \\
College education & Fraction of population with a bachelor's degree \\
Death penalty & Does/does not have the death penalty \\
Football      & Has an American football team \\
Gun laws      & Firearm carry laws \\
Medicaid      & Medicaid expansion under the ACA \\
Sales tax     & Sales tax rate \\
Same-sex marriage & Whether state law permits same-sex marriage \\
Temperature   & Annual average temperature \\
Vote 2024     & Winner of the presidential vote in 2024 \\
\end{tabular}
\end{table}

This data set provides an opportunity to showcase another application of our method, to variable selection---the identification of variables that are, or are not, particularly informative about class membership \citep{RD06,DR10,GHM11,WWM16}.  In the present case, for instance, where the classes appear to correspond to political persuasion, one might imagine that the ``Temperature'' variable, which measures annual average temperature, would not be particularly informative.  There are a number of methods for performing variable selection in mixture models.  One approach that tackles the problem head-on is to calculate the mutual information between the data and the component assignments found by the algorithm \citep{RHP22}.  Mutual information is precisely a measure of how much one random variable tells us about another, so it directly addresses the question of how much the responses tell us about class membership.  For two general random variables $x$ and~$y$, the mutual information~$I$ is defined by
\begin{equation}
I = \sum_{x,y} P(x,y) \log {P(x,y)\over P(x) P(y)},
\end{equation}
where $P(x)$, $P(y)$, and $P(x,y)$ are the marginal and joint distributions of the variables.  In the present context, where the variables are the responses~$x$ to a question~$q$ and assignments to classes~$r$, we can calculate these distributions from
\begin{align}
P(r,x) &= {1\over N} \sum_{i=1}^n \one_{z_i=r} \one_{x_{iq}=x}
        = {m_{rqx}\over N}, \\
P(r) &= {1\over N} \sum_{i=1}^n \one_{z_i=r} = {1\over n} \sum_x m_{rqx}
      = {n_r\over N}, \\
P(x) &= {1\over N} \sum_{i=1}^n \one_{x_{iq}=x} = {1\over N} \sum_r m_{rqx}
      = {n_{qx}\over N},
\end{align}
where $n_{qx}$ denotes the number of respondents (in any class) who answered question~$q$ with response~$x$, and $m_{rqx}$ and $n_r$ are as defined previously.  Now the mutual information for question~$q$ for a single class assignment is
\begin{equation}
I_q = {1\over N} \sum_{ra} m_{rqx} \log {N m_{rqx}\over n_r n_{qx}},
\end{equation}
and we average this quantity over sampled assignments to get the average mutual information.

\begin{figure}
\begin{center}
\includegraphics[width=9cm]{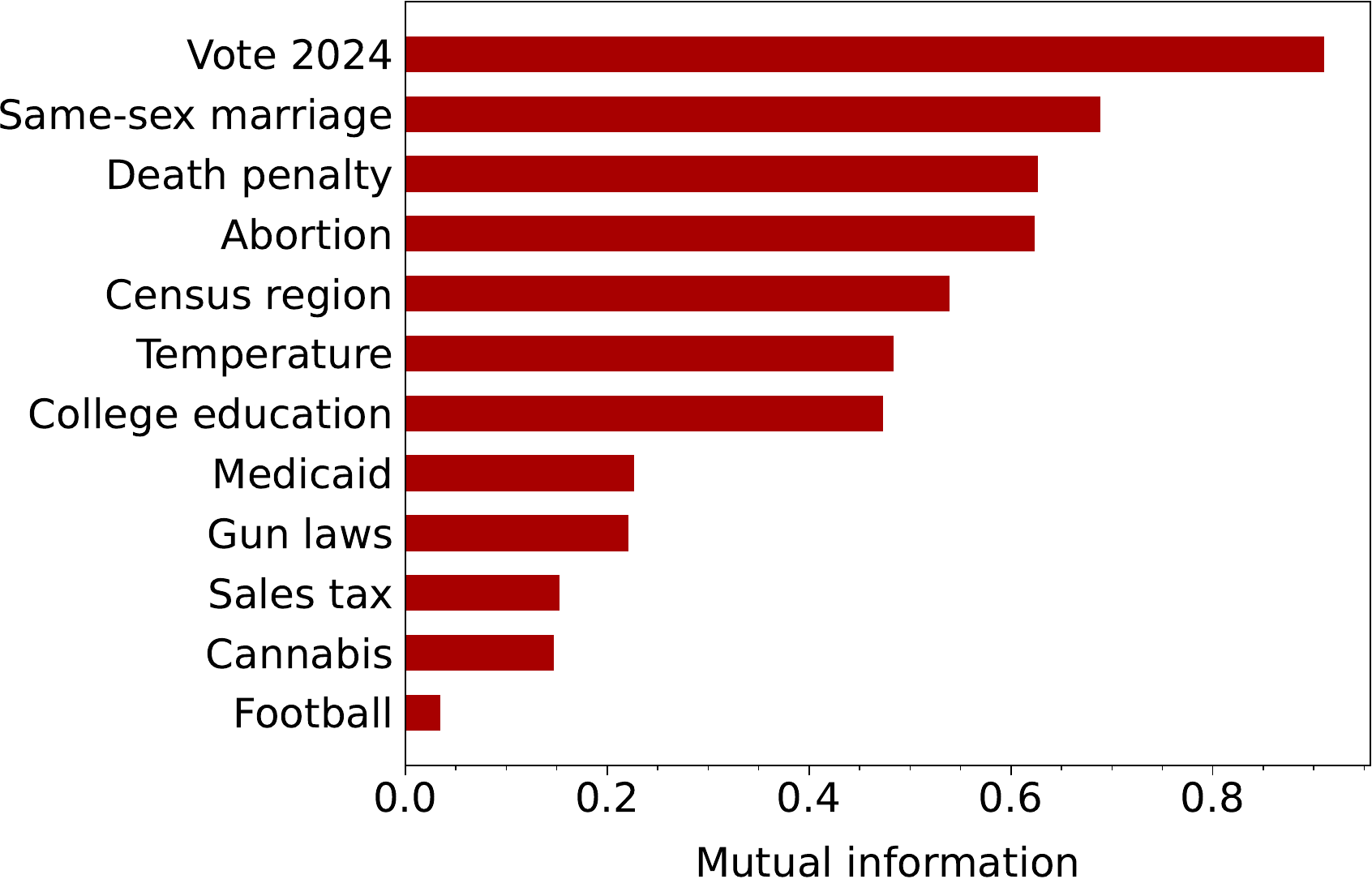}
\end{center}
\caption{Values of the mutual information (in bits) between the class assignments and the data for each question~$q$ in the 50 states data set.  High values indicate questions that are informative about class membership.}
\label{fig:mi}
\end{figure}

The mutual information is always non-negative and will take high values for questions that are informative about the classes that respondents belong to and low values for questions that are not.  Figure~\ref{fig:mi} shows the values for each of the questions/issues in our 50 states data set.  The variables with the highest values---those most informative about class membership---are the winner of the presidential vote and positions on same-sex marriage, the death penalty, and abortion.  To a lesser extent census region is also informative, presumably because left- and right-leaning states are concentrated in certain parts of the country, and, perhaps surprisingly, temperature, which we previously hypothesized would be uninformative.  We speculate that this is because temperature is acting as a proxy for geographic location.  At the other end of the scale, having a football team is uninformative, as one might expect, but so are the legality of cannabis and the amount of sales tax, issues that might seem indicative of political alignment but appear not to be in this case.

\section{Conclusions}
In this paper we have studied the use of Bayesian mixture models in model-based data clustering and described a new Monte Carlo algorithm for sampling from the integrated posterior of such models, which directly samples component assignments and the number of components, providing a way to perform both clustering and model selection in a single calculation, without the need for additional model selection steps.  Our algorithm, a form of collapsed Gibbs sampler that employs rejection-free sampling from the prior over component assignments, is substantially faster than competing algorithms in head-to-head tests, by a factor of up to fifty or more in some cases.

We have demonstrated our approach with applications to Gaussian, Poisson, and categorical models on various data sets, including extensive tests on synthetic data and a selection of real-world examples.  We find that the method is consistently able to infer the correct number of classes in the synthetic tests and provides useful results for a range of other tasks, including class assignment, consensus clustering, and variable selection.

\backmatter

\bmhead{Acknowledgments}
The author thanks Elizabeth Bruch, Carrie Ferrario, Max Jerdee, Alec Kirkley, and two anonymous referees for useful comments and feedback, and Max Jerdee, Conrad Kosowski, Gabriela Fernandes Martins, and Chethan Prakash for providing the candy dispenser data used in Section~\ref{sec:poisson}.  This work was supported in part by the US National Science Foundation under grant DMS--2404617.  Data and code are available online at \verb|https://umich.edu/~mejn/mixture|, except for previously published data, which are available at the links and references given in the text and appendices.

\setlength{\bibsep}{0.85ex}

\newpage
\begin{appendices}

\section{Proof of correctness for the\\Monte Carlo algorithm}
\label{app:mcmc}
In this appendix we prove that the distribution of samples generated by the algorithm of Section~\ref{sec:mc} converges to the integrated posterior distribution of the corresponding mixture model.

\subsection{Reformulation of the algorithm}
To prove convergence, we first give an alternate formulation of the algorithm, which would be less efficient in practice but for which the proof is simpler.  After proving the correctness of this algorithm we then demonstrate that the algorithm of Section~\ref{sec:mc}---the one we actually use---is equivalent to this alternate formulation.

A step of the alternate algorithm is as follows.

\begin{enumerate}
\setlength{\itemsep}{4pt}
\item Choose a component~$r$ uniformly at random, then choose an observation~$i$ uniformly at random from that component.
\item Remove $i$ from component~$r$.
\item If $i$ was the only member of component~$r$, delete component~$r$, relabel component~$k$ to be the new component~$r$, and decrease~$k$ by~1.  (If $r=k$ then no relabeling is necessary.)
\item Assemble a set of $2k+1$ candidate states, each of which has a weight associated with it, as follows.
\begin{enumerate}
\setlength{\itemsep}{4pt}
\item Of the $2k+1$ states, $k$~of them are states in which $i$ is placed in one of the $k$ current components~$s$.  The weights associated with each of these candidate states are
\begin{equation}
w_\mu = {P(x|k,z_{-i},z_i=s)\over P(x|k,z_{-i})},
\label{eq:w1proof}
\end{equation}
where $z_{-i}$ denotes all component assignments except for~$z_i$.
\item The remaining $k+1$ candidate states are ones that make $i$ the sole member of a new component $k+1$, swap the labels of component $k+1$ and another component $s=1\ldots k+1$, then increase $k$ by~1.  (If $s=k+1$ no swapping of labels is necessary; we simply create the new component.)  The weight associated with each of these states is
\begin{equation}
w_\mu = {k^2\over(k+1)(N-k)} {P(k+1) P(x|k+1,z_{-i},z_i=s)\over P(k) P(x|k,z_{-i})},
\label{eq:w2proof}
\end{equation}
where $P(k)$ is the prior on~$k$ as previously, $k$~is the number of components before the new component is added, and the swapping of the labels is already incorporated into the values of~$z_{-i}$.
\end{enumerate}
\item Once the set of candidate states is constructed, choose one state~$\mu$ in proportion to its weight, i.e.,~with probability
\begin{equation}
p_\mu = {w_\mu\over\sum_\nu w_\nu},
\label{eq:pimu}
\end{equation}
and update the system to that new state.
\end{enumerate}

\subsection{Ergodicity}
A Markov chain Monte Carlo algorithm converges to its target distribution if the algorithm satisfies two conditions: ergodicity and detailed balance \citep{NB99}.  The requirement of ergodicity states that the algorithm must be able to reach any state of the system from any other in a finite number of Monte Carlo steps.  In our algorithm, the states are pairs $k,z$ of number of components plus component assignment for all observations.  For these states ergodicity is trivially satisfied by the algorithm above, since our basic Monte Carlo moves can move any observation to a different component or create or delete components, and a finite series of such moves can reach a state with any value of~$k$ and component assignment~$z$.

\subsection{Detailed balance}
\label{sec:balance}
The more demanding step is the proof of detailed balance.  Detailed balance is the requirement that for any pair of states $\mu,\nu$ the average rate of transitions between them is the same in either direction in equilibrium.  If $P(\mu)$ is the equilibrium probability of being in state~$\mu$ and $P(\mu\to\nu)$ is the probability of making a transition from $\mu$ to~$\nu$, then the detailed balance condition can be written in the form
\begin{equation}
P(\mu) P(\mu\to\nu) = P(\nu) P(\nu\to\mu),
\end{equation}
or equivalently
\begin{equation}
{P(\mu\to\nu)\over P(\nu\to\mu)} = {P(\nu)\over P(\mu)}.
\label{eq:detailed}
\end{equation}

To prove that our algorithm satisfies detailed balance we must consider several cases.  First, consider moves that do not change the number of components.  There are two types of such moves.  The first are trivial moves in which we remove the last member~$i$ of a component~$r$ and delete the component, but then immediately create a new component~$s$, making $i$ its sole member.  Detailed balance is simple in this case because the reverse of such a move is the same move again, and hence the probabilities $P(\mu\to\nu)$ and $P(\nu\to\mu)$ are equal, which trivially satisfies detailed balance.

More complicated are the steps where we move an observation from one component to another without creating or deleting any components.  The probability $P(\mu\to\nu)$ for the forward move in this case is equal to the probability $1/k$ of choosing a particular component~$r$, times the probability~$1/n_r$ of picking a particular observation~$i$ from that component, times the probability~$p_\nu = w_\nu/\sum_\mu w_\mu$ of choosing the target state~$\nu$:
\begin{equation}
P(\mu\to\nu) = {1\over k}\times{1\over n_r} \times {w_\nu\over\sum_\mu w_\mu}.
\label{eq:pforward}
\end{equation}
For the backward move the corresponding probability is
\begin{equation}
P(\nu\to\mu) = {1\over k}\times{1\over n_s'} \times
               {w_\mu\over\sum_\mu w_\mu},
\label{eq:pback}
\end{equation}
where the primed variable $n_s'$ denotes the value in state~$\nu$.  The sums in the denominators of~\eqref{eq:pforward} and~\eqref{eq:pback} are over the same set of states and so have the same value, and hence, using Eq.~\eqref{eq:w1proof} for the weights, the ratio of probabilities is
\begin{equation}
{P(\mu\to\nu)\over P(\nu\to\mu)} = {n_s'\over n_r} {w_\nu\over w_\mu}
  = {n_s+1\over n_r} {P(x|k,z_{-i},z_i=s)\over P(x|k,z_{-i},z_i=r)}.
\label{eq:db1}
\end{equation}
where the remaining factors have canceled and we have made use of $n_s' = n_s + 1$.

To prove detailed balance we need to show that this ratio is equal to the ratio of posterior probabilities~$P(\nu)/P(\mu)$ as in Eq.~\eqref{eq:detailed}.  Using Eq.~\eqref{eq:posterior}, we have
\begin{equation}
P(k,z|x) = {P(k) P(x|k,z)\over P(x)} {N-1\choose k-1}^{-1}
  {\prod_r n_r!\over N!}.
\label{eq:intpost}
\end{equation}
Taking the ratio of probabilities for the states before and after the move, many factors cancel and we get
\begin{equation}
{P(\nu)\over P(\mu)}
  = {n_r'! n_s'!\over n_r! n_s!}
    {P(k)P(x|k,z_{-i},z_i=s)/P(x)\over P(k)P(x|k,z_{-i},z_i=r)/P(x)}
  = {n_s+1\over n_r} {P(x|k,z_{-i},z_i=s)\over P(x|k,z_{-i},z_i=r)},
\label{eq:pratio}
\end{equation}
where we have used $n_r' = n_r - 1$ and $n_s' = n_s+1$.

Equation~\eqref{eq:pratio} is indeed equal to Eq.~\eqref{eq:db1} and hence detailed balance is established.  The only exception is for the special case where an observation is ``moved'' to the same component it is already in, so that no change of state occurs.  However, it is easy to demonstrate that, with the choice of weights in Eq.~\eqref{eq:w1proof}, both sides of Eq.~\eqref{eq:detailed} are equal to~1 in this case, so again detailed balance is satisfied.

The last class of moves we need to consider are those that do change the number of components, either deleting a component or adding a new one.  For the purposes of our proof, and without loss of generality, let us consider the forward move $\mu\to\nu$ to be the one that deletes a component and the reverse move to be the one that adds a component.  Thus, the forward move removes the last observation~$i$ from component~$r$ and, since the component is now empty, deletes the component and replaces it with the component that was previously labeled~$k$, then decreases the value of~$k$ by one.  Then $i$ is placed in one of the other current components~$s$.  The probability of this move is equal to the probability $1/k$ of choosing component~$r$ times the probability~$p_\nu = w_\nu/\sum_\mu w_\mu$ of choosing to place $i$ in component~$s$:
\begin{equation}
P(\mu\to\nu) = {1\over k} \times {w_\nu\over\sum_\mu w_\mu},
\label{eq:delete}
\end{equation}
where $w_\nu$ is as in Eq.~\eqref{eq:w1proof} and $k$ denotes the number of components before $r$ is deleted.

The reverse move removes observation~$i$ from component~$s$ and makes it the sole member of a new component~$r$.  The probability of this move is equal to the probability $1/k'$ of choosing $s$ from the $k'$ possibilities, times the probability $1/n_s'$ of picking observation~$i$, times the probability~$p_\mu$ of choosing the move that labels the new component as component~$r$, where once again the primed variables denote values in state~$\nu$.  Thus,
\begin{equation}
P(\nu\to\mu) = {1\over k'} \times {1\over n_s'}
                \times {w_\mu\over\sum_\mu w_\mu}.
\label{eq:create}
\end{equation}

Again the sums in the denominators of~\eqref{eq:delete} and \eqref{eq:create} are equal and, using Eqs.~\eqref{eq:w1proof} and~\eqref{eq:w2proof}, the ratio of probabilities is
\begin{align}
& {P(\mu\to\nu)\over P(\nu\to\mu)}
   = {k'\over k} n_s' {w_\nu\over w_\mu} \nonumber\\
  &\qquad= {k-1\over k} (n_s+1) {P(x|k-1,z_{-i},z_i=s)\over P(x|k-1,z_{-i})}
     {(N-k+1)k\over(k-1)^2}
     {P(k-1) P(x|k-1,z_{-i})\over P(k) P(x|k,z_{-i},z_i=r)} \nonumber\\
  &\qquad= {N-k+1\over k-1} (n_s+1) {P(k-1) P(x|k-1,z_{-i},z_i=s)\over P(k) P(x|k,z_{-i},z_i=r)}.
\label{eq:db2}
\end{align}
Meanwhile, using Eq.~\eqref{eq:intpost}, the ratio of posterior probabilities for the two states, is
\begin{align}
{P(\nu)\over P(\mu)}
  &= {P(k-1)P(x|k-1,z_{-i},z_i=2)/P(x)\over P(k)P(x|k,z_{-i},z_i=r)/P(x)}
     {(k-2)!(N-k+1)!/(N-1)!\over(k-1)!(N-k)!/(N-1)!}
   {(n_s+1)!\over n_s!} \nonumber\\
  &= {N-k+1\over k-1} (n_s+1) {P(k-1) P(x|k-1,z_{-i},z_i=s)\over P(k) P(x|k,z_{-i},z_i=r)},
\end{align}
which is equal to~\eqref{eq:db2} and hence detailed balance is again established.  This completes the proof of correctness for the algorithm of this section.

Finally, we observe that the moves that create new components with labels $s=1\ldots k+1$ are all equivalent up to a label permutation and hence, if we don't care about such permutations, we can lump them all together and represent them with a single move that creates a new component~$k+1$ and carries $k+1$ times the weight, Eq.~\eqref{eq:w2proof}, of each individual move, which gives
\begin{equation}
w_\mu = {k^2\over N-k} {P(k+1) P(x|k+1,z_{-i},z_i=k+1)\over P(k) P(x|k,z_{-i})},
\end{equation}
as in Eq.~\eqref{eq:a2w2}.  This is the algorithm described in Section~\ref{sec:mc} and the one we use in our calculations.  Lumping states together like this means that when the algorithm performs a move followed by its reverse move, we may end up not in the state we started with but in an equivalent state with permuted labels.  This has no effect on the division of the observations into components or on our estimate of the number of components, but it saves us some time and complexity in the algorithm.

\subsection{Implementation}
A number of points about implementation of the algorithm are worth mentioning.  First, it is inefficient to implement it in terms of the component membership variables~$z_i$.  A better approach is to maintain instead a list in a simple array of the observations in each component, along with a record of the number~$n_r$ of members of the component.  This allows one to choose a random member of a component quickly, as required by the algorithm, and a member can be efficiently added to a component by putting it in the first free element of the array.  A member can be removed by overwriting it with the last element.

Some Monte Carlo steps also require us to renumber entire components.  Rather than renumbering each member of a component separately, which would be slow, we instead simply swap pointers to the memory locations containing the membership lists, which can be done in~$\Ord(1)$ time.

Some computational effort can also be saved by delaying the removal of a selected observation from its component.  It is straightforward to calculate the weights in Eqs.~\eqref{eq:a2w1} and~\eqref{eq:a2w2} as if the observation \emph{had} been removed, which allows us to decide which move to perform before making any updates.  With these precautions, the running time for a single Monte Carlo step of the algorithm can be reduced to~$\Ord(k)$, which is optimal since $k$ weights have to be computed for every step.

\section{Additional examples}
In this appendix we give additional example applications of the LCA algorithm from Section~\ref{sec:lca} to real-world data sets, including one data set significantly larger than any other we consider.

\subsection{Problem behaviors among teenagers}
\begin{figure}
\begin{center}
\includegraphics[width=9cm]{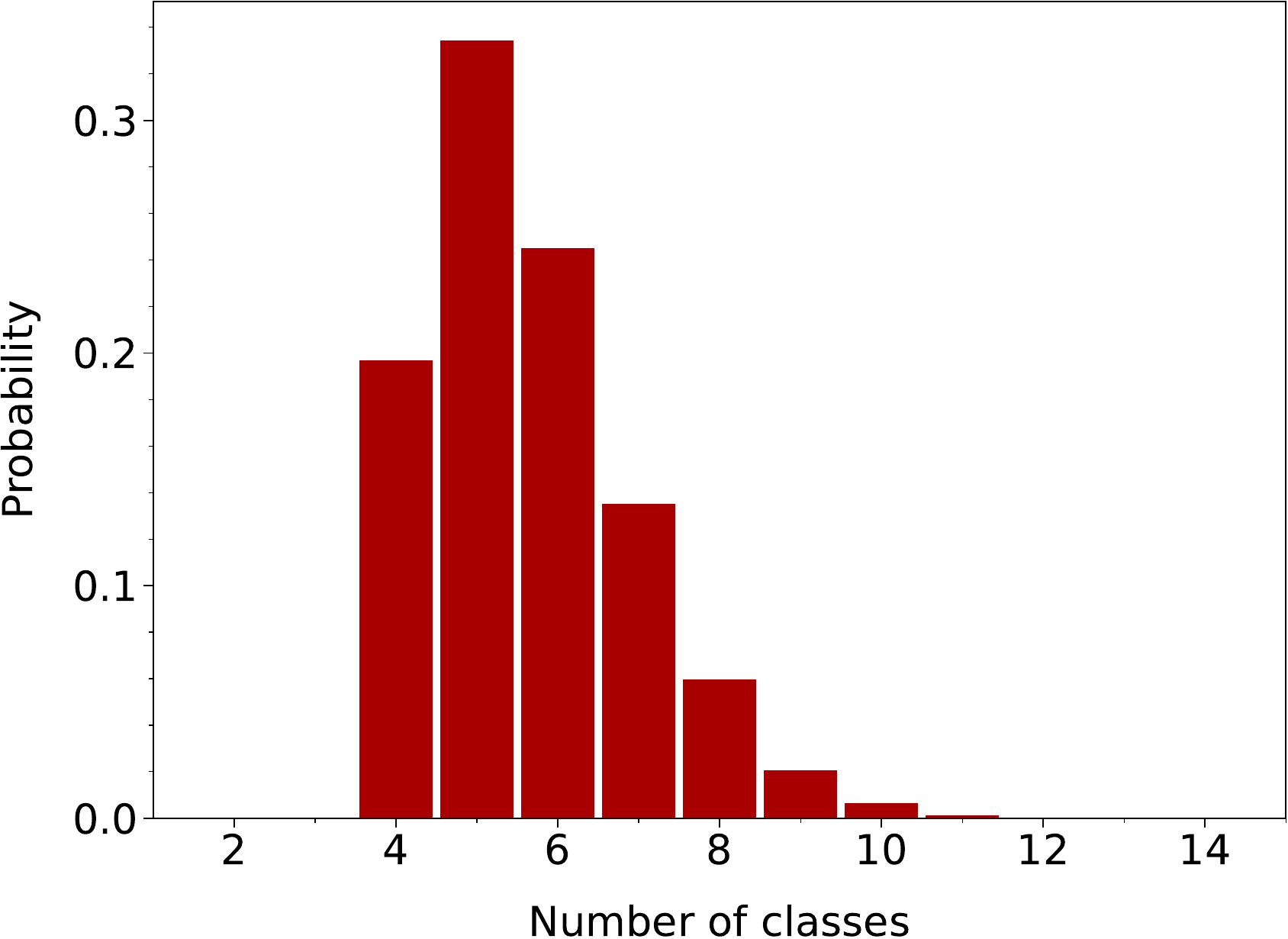}
\end{center}
\caption{Posterior distribution of the number of classes in the teenage behavior data set.}
\label{fig:behave}
\end{figure}

This example is an application to a large behavioral data set and illustrates a case in which our method gives different answers from previous approaches.  We examine a data set presented by \citet{LLSL18}, which describes survey results on the presence or absence of six problem behaviors in 6504 US teenagers.  A previous analysis of a subset of these data by \citet{CL10} found four latent classes.  Figure~\ref{fig:behave} shows the posterior distribution of the number of classes returned by our Monte Carlo method and, as we can see, four classes is a possible fit in this case, but the MAP estimate is five, and six is also more likely than four.  We find a longer run is necessary with this data set than for others we consider, in order to achieve consistent results---we ran for 250\,000 Monte Carlo sweeps, plus 25\,000 for burn-in.  Nonetheless the calculation is fast: Li~\textit{et al.}~report that their Gibbs sampling calculation took about half an hour; ours takes less than 90 seconds on roughly comparable hardware.

\subsection{Online dating}
In this example we demonstrate an application of the method to a much larger data set, which comes from the online dating service OKCupid.  Upon creating accounts on this service, users are asked a set of questions about themselves, which the service uses to match them with potential partners.  Our data set consists of answers from about 60\,000 users to 14 of these questions, which ask about gender, sexual orientation, relationship status, income, offspring, body type, diet, drinking and smoking habits, drug use, zodiac sign, religion, and whether they like cats and dogs.

\begin{figure}
\begin{center}
\includegraphics[width=9cm]{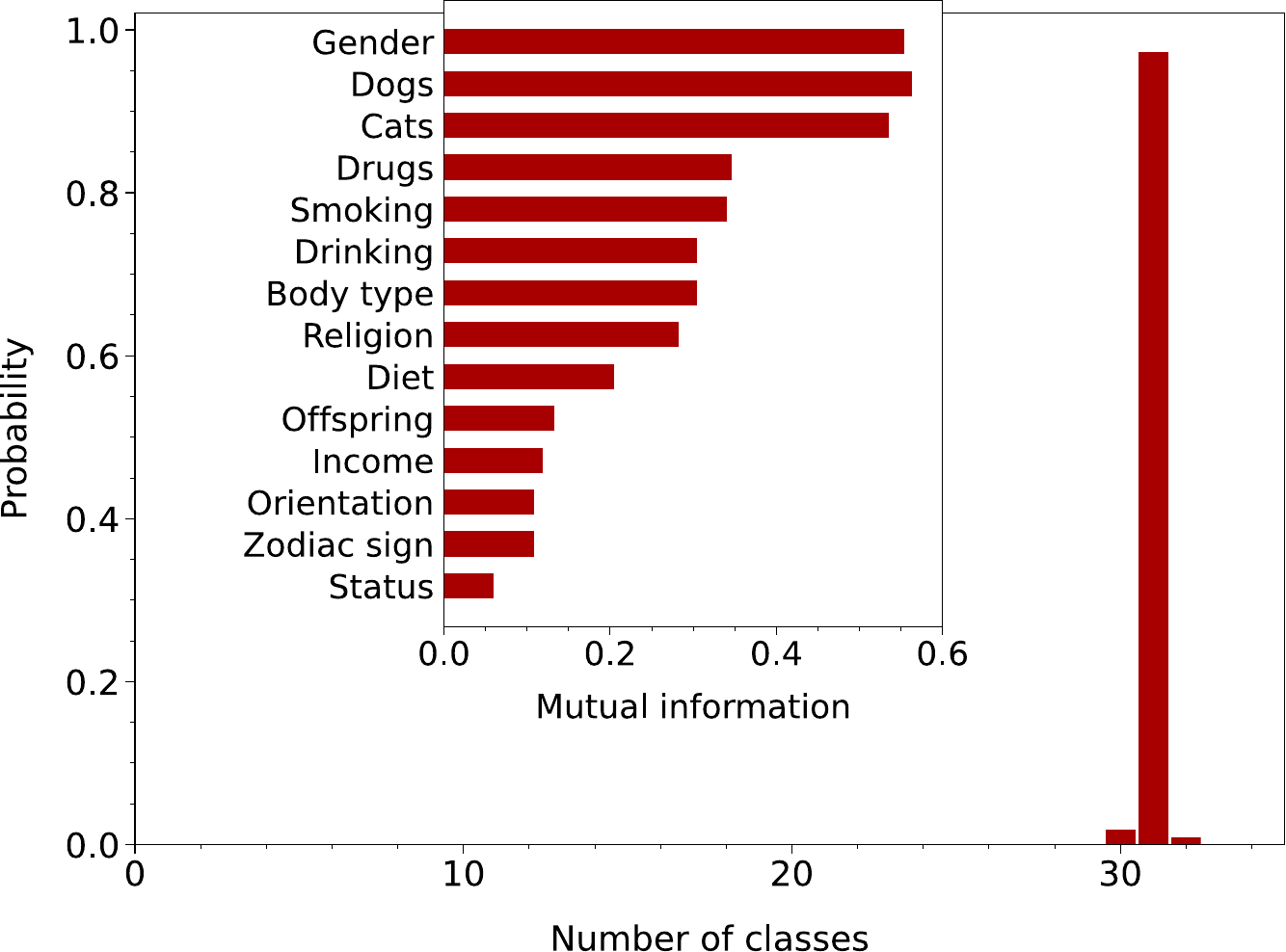}
\end{center}
\caption{Main figure: Distribution of inferred number of classes in the online dating data set.  Inset: Mutual information between data and class assignments for each of the 14 questions.}
\label{fig:dating}
\end{figure}

The analysis of this data set takes considerably longer than any of the previous ones, since one sweep now corresponds to 60\,000 Monte Carlo steps and hence takes longer to complete.  25\,000 sweeps, for example, takes about nine minutes on the author's laptop.  Figure~\ref{fig:dating} (main panel) shows the posterior distribution of the number of classes and, as we can see, this data set has a much larger number of classes than any of our others, with the distribution strongly peaked around a MAP estimate of $k=31$ classes.  Inset in the figure we show the mutual information scores for the 14 questions, calculated using the method of Section~\ref{sec:states}.  As before, these scores indicate which observations are most informative about component assignments.  The most informative observations in this case are (perhaps unsurprisingly) gender and (more surprisingly) whether the respondents like cats and dogs.  Apparently the world really is divided between cat people and dog people.  High mutual information scores also go to the questions on smoking, drinking, and drugs (which may well be correlated with one another) and to body type and religion.  At the bottom of the list are zodiac sign, which is perhaps not surprising, but also status (single, married, etc.)\ and sexual orientation.  The latter seem more surprising, but we should bear in mind that their low mutual information scores merely indicate that they are not correlated with people's answers to the other questions.  They are saying, for example, that being straight or gay doesn't affect whether you drink or like dogs.

\section{Algorithm for the model with\\general $\boldsymbol{\eta}$}
\label{app:genalpha}
The algorithm of Section~\ref{sec:mc}, which is used in our example calculations, assumes a Dirichlet-categorical prior over component assignments with concentration parameter $\eta=1$.  The methods of this paper can, however, be extended to general values of~$\eta$, for which the prior takes the form given in Eq.~\eqref{eq:nonempty}:
\begin{equation}
P(z|k,\eta) = {1\over N!} (N-k)\,\Beta(N-k,k\eta)
     \prod_{r=1}^k n_r {\Gamma(n_r+\eta-1)\over\Gamma(\eta)},
\label{eq:nonempty2}
\end{equation}
where $\Beta(x,y)$ is Euler's beta function.  The generalization makes use of a Bortz-Kalos-Lebowitz style continuous-time Monte Carlo dynamics \citep{BKL75,NB99} in which we keep track of a real-valued time variable~$t$ that measures elapsed time in arbitrary units.  For general~$\eta$, a step of the algorithm is as follows:
\begin{enumerate}
\item Define a set of weights $u_r$ for components $r=1\ldots k$ by
\begin{equation}
u_r = \biggl\lbrace \begin{array}{ll}
      1 & \qquad\mbox{if $n_r=1$,} \\
      (n_r-1)/(n_r+\eta-2) & \qquad\mbox{if $n_r>1$,}
\end{array}
\label{eq:compweights}
\end{equation}
then draw a component~$r$ at random from the categorical distribution with probabilities $u_r/\sum_s u_s$.  Simultaneously, increase the time variable according to
\begin{equation}
t \to t + {k\over\sum_s u_s}.
\label{eq:timestep}
\end{equation}
\item Choose an observation~$i$ uniformly at random from component~$r$.
\item Remove $i$ from component~$r$.
\item If $i$ was the only member of component~$r$, delete component~$r$, relabel component~$k$ to be the new component~$r$, and decrease~$k$ by~1.  (If $r=k$ then no relabeling is necessary.)
\item Assemble a set of $k+1$ candidate states~$\mu=(k,z)$, each of which has a weight~$w_\mu$ associated with it, as follows.
\begin{enumerate}
\setlength{\itemsep}{4pt}
\item Of the $k+1$ states, $k$~of them are states in which $i$ is placed in one of the $k$ current components~$s$.  Each of these candidate states has an associated weight
\begin{equation}
w_\mu = {P(x|k,z_{-i},z_i=s)\over P(x|k,z_{-i})},
\label{eq:a3w1}
\end{equation}
where as previously $z_{-i}$ denotes the set of all component assignments except for~$z_i$.
\item The last candidate state is one that makes $i$ the sole member of a new component~$k+1$ and increases $k$ by~1.  The weight associated with this state~is
\begin{equation}
w_\mu = {k(N-k-1)\,\Beta(N-k-1,(k+1)\eta)\over(N-k)\,\Beta(N-k,k\eta)}
        {P(k+1) P(x|k+1,z_{-i},z_i=k+1)\over P(k) P(x|k,z_{-i})},
\label{eq:a3w2}
\end{equation}
where $k$ is the number of components before the new component is added.
\end{enumerate}
\item Once the set of target states is assembled, choose one of them~$\mu$ with probability $p_\mu = w_\mu/\sum_\nu w_\nu$ and update the system to that new state.
\end{enumerate}
Sample states are drawn in the normal manner at uniform intervals, but using time as measured by the time variable~$t$, rather than time in Monte Carlo steps.  For the conventional choice $\eta=1$, this algorithm reduces to the algorithm of the main paper.  (We leave the demonstration as an exercise for the interested reader.)

Proof of the correctness of the algorithm follows similar lines to that of Section~\ref{sec:balance}, but allowing for the effect of the continuous time variable.  For moves that do not change the number of components, merely moving an observation from one component to another, the probability for the forward move from component~$r$ to component~$s$ is equal to the probability of choosing component~$r$, times the probability~$1/n_r$ of picking the particular observation~$i$, times the probability~$p_\nu = w_\nu/\sum_\mu w_\mu$ of choosing the candidate state~$\nu$, which gives
\begin{equation}
P(\mu\to\nu) = {u_r\over\sum_r u_r} \times {1\over n_r} \times
               {w_\nu\over\sum_\mu w_\mu}.
\end{equation}
For the backward move the probability is
\begin{equation}
P(\nu\to\mu) = {v_s\over\sum_s v_s} \times {1\over n_s'} \times
               {w_\mu\over\sum_\mu w_\mu},
\end{equation}
where $v_s$ are the weights of Eq.~\eqref{eq:compweights} for the backward move.  Bearing in mind that $n_r$ and $n_s'$ must both be greater than~1 if the number of components does not change, and making use of Eq.~\eqref{eq:compweights} for $u_r$ and~$v_s$, the ratio of forward and backward probabilities is then
\begin{align}
{P(\mu\to\nu)\over P(\nu\to\mu)}
  &= {u_r/\sum_r u_r\over v_s/\sum_s v_s} {n_s'\over n_r}\, {w_\nu\over w_\mu}
     \nonumber\\
  &= {(n_r-1)/n_r\over n_s/(n_s+1)} 
     \biggl( {n_s+\eta-1\over n_r+\eta-2} \biggr)
     {P(x|k,z_{-i},z_i=s)\over P(x|k,z_{-i},z_i=r)}\,{\sum_s v_s\over\sum_r u_r}.
\label{eq:gendet1}
\end{align}

Now we compare this to the ratio of the desired equilibrium probabilities of the corresponding states.  We have
\begin{equation}
P(k,z|x,\eta) = {P(x|k,z) P(z|k,\eta) P(k)\over P(x)},
\end{equation}
and hence, using~\eqref{eq:nonempty2}, we have
\begin{align}
{P(\nu)\over P(\mu)}
  &= {n_r' \Gamma(n_r'+\eta-1) n_s' \Gamma(n_s'+\eta-1) P(x|k,z_{-i},z_i=s) P(k)/P(x)
      \over
      n_r \Gamma(n_r+\eta-1) n_s \Gamma(n_s+\eta-1) P(x|k,z_{-i},z_i=r) P(k)/P(x)}
      \nonumber\\
  &= {(n_r-1)/n_r\over n_s/(n_s+1)}\, \biggl( {n_s+\eta-1\over n_r+\eta-2}
     \biggr) {P(x|k,z_{-i},z_i=s)\over P(x|k,z_{-i},z_i=r)}.
\end{align}
Substituting this expression into Eq.~\eqref{eq:gendet1} we have
\begin{equation}
{P(\mu\to\nu)\over P(\nu\to\mu)}
  = {P(\nu)\,(1/k)\sum_s v_s\over P(\mu)\,(1/k)\sum_r u_r},
\end{equation}
which is similar to the standard detailed balance formula, Eq.~\eqref{eq:detailed}, except for the factors of $(1/k)\sum_r u_r$ and $(1/k)\sum_s v_s$.  This means that each state~$\mu$ will appear not with its normal probability~$P(\mu)$ but with modified probability $P(\mu) (1/k)\sum_r u_r$.  However, the probability of \emph{sampling} from state~$\mu$ is multiplied by the amount of time for which the system remains in that state, which is $k/\sum_r u_r$, following Eq.~\eqref{eq:timestep}.  So the probability of sampling is precisely proportional to~$P(\mu)$, as required.

For Monte Carlo steps that change the number of components, we can once more, without loss of generality, consider the forward move $\mu\to\nu$ to be the one that removes the last observation~$i$ from a component~$r$ and then deletes the empty component, and the reverse move to be the one that creates a new component.  The probability of the forward move is equal to the probability of choosing component~$r$ times the probability $p_\nu = w_\nu/\sum_\mu w_\mu$ of placing observation~$i$ in a particular new component~$s$:
\begin{equation}
P(\mu\to\nu) = {u_r\over\sum_r u_r} \times {w_\nu\over\sum_\mu w_\mu}.
\end{equation}

The backward move removes observation~$i$ from component~$s$ and makes it the sole member of a new component~$r$.  The probability for this move is equal to the probability of choosing component~$s$, times the probability~$1/n_s'$ of picking observation~$i$, times the probability~$p_\mu$ of choosing the move that creates the new component.  As with the proof in Section~\ref{app:mcmc}, we will initially assume separate weights for creating new components with each possible component label~$1\ldots k'+1$ and values
\begin{align}
w_\mu &= {k'(N-k'-1)\,\Beta(N-k'-1,(k'+1)\eta)\over(k'+1)(N-k')\,\Beta(N-k',k'\eta)}
        {P(k'+1) P(x|k'+1,z_{-i},z_i=r)\over P(k') P(x|k',z_{-i})} \nonumber\\
  &= {(k-1)(N-k)\,\Beta(N-k,k\eta)\over k(N-k+1)\,\Beta(N-k+1,(k-1)\eta)}
        {P(k) P(x|k,z_{-i},z_i=r)\over P(k-1) P(x|k-1,z_{-i})}.
\label{eq:wmugen}
\end{align}
Then we have
\begin{equation}
P(\nu\to\mu) = {v_s\over\sum_s v_s} \times {1\over n_s'} \times
               {w_\mu\over\sum_\mu w_\mu},
\end{equation}
and the ratio of probabilities is
\begin{align}
{P(\mu\to\nu)\over P(\nu\to\mu)}
  &= {u_r/\sum_r u_r\over v_s/\sum_s v_s} n_s' {w_\nu\over w_\mu}
   = {n_s+\eta-1\over n_s} {\sum_s v_s\over\sum_r v_r} (n_s+1) \nonumber\\
  &\quad\times
     {k(N-k+1)\,\Beta(N-k+1,(k-1)\eta)\over(k-1)(N-k)\,\Beta(N-k,k\eta)}
     {P(k-1) P(x|k-1,z_{-i},z_i=s)\over P(k) P(x|k,z_{-i},z_i=r)}.
\label{eq:gendet2}
\end{align}

Meanwhile, the ratio of equilibrium probabilities for states~$\mu$ and~$\nu$ is
\begin{align}
{P(\nu)\over P(\mu)}
  &= {(N-k+1)\,\Beta(N-k+1,(k-1)\eta)\over(N-k)\,\Beta(N-k,k\eta)}
     {(n_s+1)\Gamma(n_s+\eta)\over\Gamma(\eta)}
     {\Gamma(\eta)^2\over \Gamma(\eta) n_s \Gamma(n_s+\eta-1)} \nonumber\\
  &\qquad\times {P(x|k-1,z_{-i},z_i=s) P(k-1)/P(x)\over P(x|k,z_{-i},z_i=r) P(k)/P(x)} \nonumber\\
  &= (n_s+\eta-1) {n_s+1\over n_s}
     {(N-k+1)\,\Beta(N-k+1,(k-1)\eta)\over(N-k)\,\Beta(N-k,k\eta)} \nonumber\\
  &\qquad\times {P(k-1) P(x|k-1,z_{-i},z_i=s)\over P(k) P(x|k,z_{-i},z_i=r)}.
\end{align}
Comparing with Eq.~\eqref{eq:gendet2}, we find that
\begin{equation}
{P(\nu)\over P(\mu)} = {P(\nu) [1/(k-1)] \sum_s v_s\over
                        P(\mu) (1/k) \sum_r u_r}.
\end{equation}
Hence, once again, we have a modified detailed balance condition that implies that each state~$\mu$ will appear with probability $P(\mu) (1/k) \sum_r u_r$.  But the probability of \emph{sampling} each state is multiplied by the length of time for which the system remains in that state, which is $k/\sum_r u_r$ following Eq.~\eqref{eq:timestep}, and hence the probability of sampling is exactly proportional to~$P(\mu)$, as required.

The final step of the proof, as in Section~\ref{app:mcmc}, is to combine the $k+1$ moves that create new components numbered $1\ldots k+1$ into a single move that creates a component numbered $k+1$, with $k+1$ times the probability.  The weight for this move is given by Eq.~\eqref{eq:wmugen} times $k'+1$, which gives the expression in Eq.~\eqref{eq:a3w2}.  As previously, this can result in a permutation of the labels of the components, but it has no effect on the actual partition of the observations, since the component labels are arbitrary.  This completes the proof of correctness for the generalized algorithm.

Implementation of the algorithm is straightforward and the slightly higher level of complexity compared with the case for $\eta=1$ does not impact performance significantly.  In particular, note that one can calculate in advance tables of $n/(n+\eta-1)$ and $\Beta(N-k,k\eta)$ for $n,k = 1\ldots N$ and then use them to evaluate Eqs.~\eqref{eq:compweights} and~\eqref{eq:a3w2} with little performance penalty.

\section{Data sets}
\label{app:data}
In this appendix we describe the data sets used in our examples.

\subsection{Epileptic seizures}
These data come from a clinical trial of medications for pediatric epilepsy reported by \citet{WPCL96}.  The data describe the experiences of a single epileptic child participant in the trial over a 140-day period.  The participant was observed under baseline conditions for 28 days, then received monthly infusions of intravenous gammaglobulin, as a potential treatment, for the remainder of the observation period, and the data set consists of the number of seizure episodes experienced on each day of the trial, as recorded in a diary by the child's parent.  The complete data set is included in the paper by Wang~\textit{et al.}

\subsection{Candy dispenser}
This data set comes from an automated candy dispenser in the reception area outside the author's office, which is stocked with M\&Ms brand candy.  The dispenser uses a motorized corkscrew mechanism to dispense a handful of M\&Ms upon the push of a button.  The data set records the number of candies dispensed on 857 pushes of the button and was gathered by Max Jerdee, Conrad Kosowski, Gabriela Fernandes Martins, and Chethan Prakash.  The measurements were near-consecutive, but four measurements had to be discarded because of missing data, leaving a total of 853 for analysis.  A~copy of the data set is available for download on the web at \verb|https://umich.edu/~mejn/mixture|.

\begin{table}
\caption{The 32 questions selected from the CBS/New York Times poll for the data set used in this paper, listed in the order in which they were asked on the survey.}
\label{tab:cbs}
\setlength{\tabcolsep}{8pt}
\begin{tabular}{llc}
Item & Question & Responses \\
\hline
\verb|sex|  & Respondent's sex       & 2 \\
\verb|cenr| & Census region          & 4 \\
\verb|q1|   & Obama job approval     & 3 \\
\verb|q2|   & Right direction/wrong track & 3 \\
\verb|q6|   & Congress job approval  & 3 \\
\verb|q13|  & Rate national economy  & 5 \\
\verb|reg|  & Registered voter       & 3 \\
\verb|q66|  & Spending cuts vs.\ jobs & 4 \\
\verb|q67|  & Payroll tax cut        & 3 \\
\verb|q69|  & Spending on infrastructure & 3 \\
\verb|q70|  & Small business tax cut & 3 \\
\verb|q76|  & Tax on wealthy         & 3 \\
\verb|q78|  & Economic liberal/conservative & 6 \\
\verb|q79|  & Social liberal/conservative & 6 \\
\verb|q80|  & Social security/Medicare & 3 \\
\verb|q84|  & Repeal health care law & 4 \\
\verb|q86|  & Death penalty          & 3 \\
\verb|q88|  & Global warming         & 6 \\
\verb|q89|  & Same-sex marriage      & 4 \\
\verb|q90|  & Abortion               & 4 \\
\verb|q92|  & Illegal immigration    & 4 \\
\verb|prty| & Party ID               & 4 \\
\verb|q106| & Employment status      & 5 \\
\verb|vt08| & Voted for in 2008      & 6 \\
\verb|evan| & Evangelical            & 3 \\
\verb|reli| & Religion               & 7 \\
\verb|marr| & Marital status         & 6 \\
\verb|agea| & Age group              & 5 \\
\verb|educ| & Education              & 6 \\
\verb|hisp| & Hispanic               & 3 \\
\verb|race| & Race                   & 5 \\
\verb|inca| & Income group           & 6 \\
\end{tabular}
\end{table}

\subsection{CBS/New York Times poll}
These data come from an opinion poll, fielded jointly by the CBS television network and the New York Times newspaper, of 1566 members of the American public during September 2011, and focusing on a range of topics of then-current interest.  The data set we examine contains responses for all participants but only a 32-question subset of the questions asked, as listed in Table~\ref{tab:cbs}.  The data were provided by the Inter-university Consortium for Political and Social Research (ICPSR) and are freely available from the ICPSR web site, file number~34458, at \verb|https://doi.org/10.3886/ICPSR34458.v1|.

\subsection{Alzheimer's diagnostic survey}
This data set, which comes from \citet{Walsh06}, records the presence or absence of six symptoms of potential Alzheimer's disease in 240 patients at the Mercer's Institute national memory clinic in Dublin, Ireland, as reported by the patients' primary caregivers on the occasion of the patients' first visit to the clinic.  The patients were preselected as having suspected Alzheimer's disease or other age-related cognitive decline, but only mild symptoms, so that a positive diagnosis was not unlikely but also not guaranteed.  The symptoms identified were: activity disturbance, affective disorder, aggression, agitation, diurnal rhythm disturbance, and hallucinations.  The survey, a standard diagnostic instrument called the Behave-AD questionnaire, returns information on both the presence and the severity of each symptom, but Walsh converted the data to dichotomous presence/absence variables and this is the version we analyze.  The data set is included in full in a table within the paper by Walsh.

\begin{table}
\caption{Summary of the 12 variables used in the 50 states data set.  ``College education'' measures the fraction of the population over the age of 25 with a bachelor's degree.  ``Football'' refers to whether the state is home to an American football team in the US National Football League.  ``Gun laws'' distinguishes constitutional carry, concealed carry plus unlicensed open carry, concealed carry plus licensed open carry, and concealed carry only.  ``Medicaid'' refers to adoption of the Medicaid expansion under the Affordable Care Act of 2010, at the time of writing of this paper.  ``Same-sex marriage'' refers to state statutes and constitutions that contain, or do not contain, wording forbidding same-sex marriage.  Actual same-sex marriage has been legal nationwide in the US since 2015 under federal law, which preempts state laws.}
\label{tab:statesfull}
\begin{tabular}{ll}
Variable & Values \\
\hline
Abortion      & Legal, Limited, Illegal \\
Cannabis      & Legal, Limited, Illegal \\
Census region & Northeast, Midwest, South, West \\
College education & Fraction with bachelor's: below $30\%$, 30--35\%, 35--40\%, above $40\%$ \\
Death penalty & Yes, not enforced, no \\
Football      & Has an NFL team, does not have an NFL team \\
Gun laws      & Yes, unlicensed open carry, licensed open carry, concealed carry only \\
Medicaid      & Expanded under the ACA, not expanded \\
Sales tax     & Below 1.75\%, 1.75--7.07\%, above 7.07\% \\
Same-sex marriage & Banned by statute or constitution, not banned \\
Temperature   & Average temperature: below $5^\circ$C, 5--$15^\circ$C, above $15^\circ$C \\
Vote 2024     & Harris, Trump \\
\end{tabular}
\end{table}

\subsection{50 states data set}
\label{app:states}
This data set, which was assembled by the current author from online sources for the purposes of this study, records 12 traits for each of the 50 United States (but excluding the District of Columbia, Puerto Rico, and other non-state territories).  The 12 traits are listed in Table~\ref{tab:statesfull} and further details are given in the table caption.  The full data set is available for download from \verb|https://umich.edu/~mejn/mixture|.

\subsection{Problem behaviors among teenagers}
This data set comes from \citet{LLSL18} and describes self-reported incidence of problem behaviors by  6504 teenagers in the United States during the 1990s.  The behaviors probed were: lying to parents, unruly public behavior, damaging property, stealing something worth less than US\$50, stealing from a store, and taking part in a fight.  Each behavior is recorded simply as present or absent, so there are only two possible responses for each survey item.  The data set is included in full in a table within the paper by Li~\textit{et al.}

\begin{table}
\caption{Summary of the 14 variables in the online dating data set.  Missing data is also coded as an additional possible answer for each question.}
\label{tab:dating}
\begin{tabular}{lp{10.8cm}}
Variable & Values \\
\hline
Gender      & Male, female \\
Orientation & Straight, gay, bisexual \\
Status      & Single, available, seeing someone, married \\
Income      & Less than \$20k, over \$20k, \$30k, \$40k, \$50k, \$60k, \$70k, \$80k, \$100k, \$150k, \$250k, \$500k, \$1m \\
Offspring   & Has children, doesn't have children, wants children, might want children, doesn't want children \\
Body type   & Skinny, thin, average, fit, athletic, a little extra, curvy, full figured, jacked, used up, overweight, rather not say \\
Diet        & Anything, vegetarian, vegan, kosher, halal, other \\
Drinking    & Not at all, rarely, socially, often, very often, desperately \\
Smoking     & No, yes, sometimes, when drinking, trying to quit \\
Drug use    & Never, sometimes, often \\
Zodiac sign & Aries, Taurus, Gemini, Cancer, Leo, Virgo, Libra, Scorpio, Sagittarius, Capricorn, Aquarius, Pisces \\
Religion    & Atheism, agnosticism, Christianity, Catholicism, Judaism, Islam, Hinduism, Buddhism, other \\
Cats        & Has cat(s), likes cats, dislikes cats \\
Dogs        & Has dog(s), likes dogs, dislikes dogs \\
\end{tabular}
\end{table}

\subsection{Online dating}
This data set contains dating profiles for 59\,946 users of the online dating service OK\-Cupid, as posted at \texttt{https://www.kaggle.com/datasets/subhamyadav580/ dating-site} by Shubham Yadav.  The full data set contains both textual descriptions contributed by the users and answers to multiple-choice profile questions, but we focus on the latter only, and on 14 questions in particular, which are listed in Table~\ref{tab:dating}.  We include a ``missing data'' option as an additional possible response for each question, since significant numbers of people did not answer all questions.

\end{appendices}


\begin{thebibliography}{63}
\expandafter\ifx\csname natexlab\endcsname\relax\def\natexlab#1{#1}\fi
\expandafter\ifx\csname url\endcsname\relax
  \def\url#1{\texttt{#1}}\fi
\expandafter\ifx\csname urlprefix\endcsname\relax\def\urlprefix{URL }\fi

\bibitem[{Arthur(2024)}]{Arthur24}
Arthur, R., 2024.
\newblock Exploring network structure with the density of states.
\newblock Preprint arxiv:2410.18253.

\bibitem[{Banfield and Raftery(1993)}]{BR93}
Banfield, J.~D. and Raftery, A.~E., 1993.
\newblock Model-based {G}aussian and non-{G}aussian clustering.
\newblock \textit{Biometrics} \textbf{49}, 803--821.

\bibitem[{Bensmail \textit{et~al.}(1997)Bensmail, Celeux, Raftery, and
  Robert}]{BCRR97}
Bensmail, H., Celeux, G., Raftery, A.~E., and Robert, C.~P., 1997.
\newblock Inference in model-based cluster analysis.
\newblock \textit{Statistics and Computing} \textbf{7}, 1--10.

\bibitem[{Beraha \textit{et~al.}(2025)Beraha, Guindani, Gianella, and
  Guglielmi}]{BGGG25}
Beraha, M., Guindani, B., Gianella, M., and Guglielmi, A., 2025.
\newblock {BayesMix}: {B}ayesian mixture models in {C++}.
\newblock \textit{Journal of Statistical Software} \textbf{112}, 9.

\bibitem[{Biernacki \textit{et~al.}(2010)Biernacki, Celeux, and
  Govaert}]{BCG10}
Biernacki, C., Celeux, G., and Govaert, G., 2010.
\newblock Exact and {M}onte {C}arlo calculations of integrated likelihoods for
  the latent class model.
\newblock \textit{Journal of Statistical Planning and Inference} \textbf{140},
  2991--3002.

\bibitem[{B{\"o}hning and Seidel(2003)}]{BS03b}
B{\"o}hning, D. and Seidel, W., 2003.
\newblock Recent developments in mixture models.
\newblock \textit{Computational Statistics and Data Analysis} \textbf{41},
  349--357.

\bibitem[{Bortz \textit{et~al.}(1975)Bortz, Kalos, and Lebowitz}]{BKL75}
Bortz, A.~B., Kalos, M.~H., and Lebowitz, J.~L., 1975.
\newblock A new algorithm for {M}onte {C}arlo simulation of ising spin systems.
\newblock \textit{Journal of Computational Physics} \textbf{17}, 10--18.

\bibitem[{Bryant(2003)}]{Bryant03}
Bryant, D., 2003.
\newblock A classification of consensus methods for phylogenies.
\newblock In M.~Janowitz, F.-J. Lapointe, F.~R. McMorris, B.~Mirkin, and
  F.~Roberts (eds.), \textit{BioConsensus}, no.~61 in DIMACS Series in Discrete
  Mathematics and Theoretical Computer Science, pp. 163--184. American
  Mathematical Society, Providence, RI.

\bibitem[{Collins and Lanza(2010)}]{CL10}
Collins, L.~M. and Lanza, S.~T., 2010.
\newblock \textit{Latent Class and Latent Transition Analysis}.
\newblock John Wiley and Sons, Hoboken, NJ.

\bibitem[{Das(2014)}]{Das14}
Das, R., 2014.
\newblock Collapsed {G}ibbs sampler for {D}irichlet process {G}aussian mixture
  models.
\newblock Technical report, School of Computer Science, Carnegie Mellon
  University.

\bibitem[{Dean and Raftery(2009)}]{DR10}
Dean, N. and Raftery, A.~E., 2009.
\newblock Latent class analysis variable selection.
\newblock \textit{Ann. Inst. Stat. Math.} \textbf{62}, 11--35.

\bibitem[{Dempster \textit{et~al.}(1977)Dempster, Laird, and Rubin}]{DLR77}
Dempster, A.~P., Laird, N.~M., and Rubin, D.~B., 1977.
\newblock Maximum likelihood from incomplete data via the {EM} algorithm.
\newblock \textit{J. R. Statist. Soc. B} \textbf{39}, 185--197.

\bibitem[{Everitt(1988)}]{Everitt88}
Everitt, B.~S., 1988.
\newblock A {M}onte {C}arlo investigation of the likelihood ratio test for
  number of classes in latent class analysis.
\newblock \textit{Multivariate Behavioral Research} \textbf{23}, 531--538.

\bibitem[{Fr{\"u}hwirth-Schnatter(2006)}]{Fruhwirth06}
Fr{\"u}hwirth-Schnatter, S., 2006.
\newblock \textit{Finite Mixture and Markov Switching Models}.
\newblock Springer, New York.

\bibitem[{Fr{\"u}hwirth-Schnatter and Malsiner-Walli(2019)}]{FM19}
Fr{\"u}hwirth-Schnatter, S. and Malsiner-Walli, G., 2019.
\newblock From here to infinity: Sparse finite versus {D}irichlet process
  mixtures in model-based clustering.
\newblock \textit{Advances in Data Analysis and Classification} \textbf{13},
  33--64.

\bibitem[{Gelfand(2000)}]{Gelfand00}
Gelfand, A.~E., 2000.
\newblock Gibbs sampling.
\newblock \textit{Journal of the American Statistical Association} \textbf{95},
  1300--1304.

\bibitem[{Gelman \textit{et~al.}(2013)Gelman, Carlin, Stern, Dunson, Vehtari,
  and Rubin}]{Gelman13}
Gelman, A., Carlin, J.~B., Stern, H.~S., Dunson, D.~B., Vehtari, A., and Rubin,
  D.~B., 2013.
\newblock \textit{Bayesian Data Analysis}.
\newblock Chapman and Hall/CRC Press, Boca Raton, FL, 3rd ed.

\bibitem[{Ghosh \textit{et~al.}(2011)Ghosh, Herring, and Siega-Riz}]{GHM11}
Ghosh, J., Herring, A.~H., and Siega-Riz, A.~M., 2011.
\newblock Bayesian variable selection for latent class models.
\newblock \textit{Biometrics} \textbf{67}, 917--925.

\bibitem[{Goder and Filkov(2008)}]{GF08}
Goder, A. and Filkov, V., 2008.
\newblock Consensus clustering algorithms: Comparison and refinement.
\newblock In \textit{Proceedings of the 10th Workshop on Algorithm Engineering
  and Experiments (ALENEX)}, pp. 109--117. Society of Industrial and Applied
  Mathematics.

\bibitem[{Good \textit{et~al.}(2010)Good, de~Montjoye, and Clauset}]{GDC10}
Good, B.~H., de~Montjoye, Y.-A., and Clauset, A., 2010.
\newblock Performance of modularity maximization in practical contexts.
\newblock \textit{Phys. Rev. E} \textbf{81}, 046106.

\bibitem[{Goodman(1974)}]{Goodman74}
Goodman, L.~A., 1974.
\newblock Exploratory latent structure analysis using both identifiable and
  unidentifiable models.
\newblock \textit{Biometrika} \textbf{61}, 215--231.

\bibitem[{Gormley \textit{et~al.}(2022)Gormley, Murphy, and Raftery}]{GMR22}
Gormley, I.~C., Murphy, T.~B., and Raftery, A.~E., 2022.
\newblock Model-based clustering.
\newblock \textit{Annual Review of Statistics and Its Application} \textbf{10},
  573--595.

\bibitem[{Green(1995)}]{Green95}
Green, P., 1995.
\newblock Reversible jump {M}arkov chain {M}onte {C}arlo computation and
  {B}ayesian model determination.
\newblock \textit{Biometrika} \textbf{82}, 711--732.

\bibitem[{Hjort \textit{et~al.}(2010)Hjort, Holmes, M{\"u}ller, and
  Walker}]{HHMW10}
Hjort, N.~L., Holmes, C., M{\"u}ller, P., and Walker, S.~G. (eds.), 2010.
\newblock \textit{Bayesian Nonparametrics}.
\newblock No.~28 in Cambridge Series in Statistical and Probabilistic
  Mathematics. Cambridge University Press, Cambridge.

\bibitem[{Hoijtink(2010)}]{Hoijtink10}
Hoijtink, H., 2010.
\newblock Confirmatory latent class analysis: Model selection using {B}ayes
  factors and (pseudo) likelihood ratio statistics.
\newblock \textit{Multivariate Behavioral Research} \textbf{36}, 563--588.

\bibitem[{Jain and Neal(2004)}]{JN04}
Jain, S. and Neal, R.~M., 2004.
\newblock A split-merge {M}arkov chain {M}onte {C}arlo procedure for the
  {D}irichlet process mixture model.
\newblock \textit{Journal of Computational and Graphical Statistics}
  \textbf{13}, 158--182.

\bibitem[{Khoufache \textit{et~al.}(2023)Khoufache, Lebbah, Azzag, Goffinet,
  and Bouchaffra}]{Khoufache23}
Khoufache, R., Lebbah, M., Azzag, H., Goffinet, E., and Bouchaffra, D., 2023.
\newblock Distributed collapsed {G}ibbs sampler for {D}irichlet process mixture
  models in federated learning.
\newblock Preprint arxiv:2312.11169.

\bibitem[{Kirkley and Newman(2022)}]{KN22}
Kirkley, A. and Newman, M. E.~J., 2022.
\newblock Representative community divisions of networks.
\newblock \textit{Communications Physics} \textbf{5}, 40.

\bibitem[{Lancichinetti and Fortunato(2012)}]{LF12}
Lancichinetti, A. and Fortunato, S., 2012.
\newblock Consensus clustering in complex networks.
\newblock \textit{Scientific Reports} \textbf{2}, 336.

\bibitem[{Li \textit{et~al.}(2018)Li, Lord-Bessen, Shiyko, and Loeb}]{LLSL18}
Li, Y., Lord-Bessen, J., Shiyko, M., and Loeb, R., 2018.
\newblock Bayesian latent class analysis tutorial.
\newblock \textit{Multivariate Behavioral Research} \textbf{53}, 430--451.

\bibitem[{Lin and Dayton(1997)}]{LD97}
Lin, T.~H. and Dayton, C.~M., 1997.
\newblock Model selection information criteria for non-nested latent class
  models.
\newblock \textit{Journal of Educational and Behavioral Statistics}
  \textbf{22}, 249--264.

\bibitem[{Marin \textit{et~al.}(2005)Marin, Mengersen, and Robert}]{MMR05}
Marin, J.-M., Mengersen, K., and Robert, C.~P., 2005.
\newblock Bayesian modelling and inference on mixtures of distributions.
\newblock \textit{Handbook of Statistics} \textbf{25}, 459--507.

\bibitem[{McCutcheon(1987)}]{McCutcheon87}
McCutcheon, A.~C., 1987.
\newblock \textit{Latent Class Analysis}.
\newblock Sage, Beverly Hill, CA.

\bibitem[{McLachlan and Peel(2000)}]{MP00}
McLachlan, G.~J. and Peel, D., 2000.
\newblock \textit{Finite Mixture Models}.
\newblock Wiley, New York.

\bibitem[{McLachlan and Rathnayake(2014)}]{MR14}
McLachlan, G.~J. and Rathnayake, S., 2014.
\newblock On the number of components in a {G}aussian mixture model.
\newblock \textit{Wiley Interdisciplinary Reviews: Data Mining and Knowledge
  Discovery} \textbf{4}, 341--355.

\bibitem[{McLachlan \textit{et~al.}(2019)McLachlan, Lee, and
  Rathnayake}]{MLR19}
McLachlan, G.~J., Lee, S.~X., and Rathnayake, S.~I., 2019.
\newblock Finite mixture models.
\newblock \textit{Annual Review of Statistics and Its Application} \textbf{6},
  355--378.

\bibitem[{Miller and Harrison(2018)}]{MH18}
Miller, J.~W. and Harrison, M.~T., 2018.
\newblock Mixture models with a prior on the number of components.
\newblock \textit{J. Amer. Stat. Assoc.} \textbf{113}, 340--356.

\bibitem[{Monti \textit{et~al.}(2003)Monti, Tamayo, Mesirov, and
  Golub}]{MPMG03}
Monti, S., Tamayo, P., Mesirov, J., and Golub, T., 2003.
\newblock Consensus clustering: A resampling-based method for class discovery
  and visualization of gene expression microarray data.
\newblock \textit{Machine Learning} \textbf{52}, 91--118.

\bibitem[{Moran \textit{et~al.}(2004)Moran, Walsh, Lynch, Coen, Coakley, and
  Lawlor}]{Moran04}
Moran, M., Walsh, C., Lynch, A., Coen, R.~F., Coakley, D., and Lawlor, B.~A.,
  2004.
\newblock Syndromes of behavioural and psychological symptoms in mild
  {A}lzheimer’s disease.
\newblock \textit{International Journal of Geriatric Psychiatry} \textbf{19},
  359--364.

\bibitem[{Nasserinejad \textit{et~al.}(2017)Nasserinejad, van Rosmalen,
  de~Kort, and Lesaffre}]{NVDL17}
Nasserinejad, K., van Rosmalen, J., de~Kort, W., and Lesaffre, E., 2017.
\newblock Comparison of criteria for choosing the number of classes in
  {B}ayesian finite mixture models.
\newblock \textit{PLOS One} \textbf{12}, e0168838.

\bibitem[{Neal(2000)}]{Neal00}
Neal, R.~M., 2000.
\newblock Markov chain sampling methods for {D}irichlet process mixture models.
\newblock \textit{Journal of Computational and Graphical Statistics}
  \textbf{9}, 249--265.

\bibitem[{Newman and Barkema(1999)}]{NB99}
Newman, M. E.~J. and Barkema, G.~T., 1999.
\newblock \textit{Monte Carlo Methods in Statistical Physics}.
\newblock Oxford University Press, Oxford.

\bibitem[{Newman and Reinert(2016)}]{NR16}
Newman, M. E.~J. and Reinert, G., 2016.
\newblock Estimating the number of communities in a network.
\newblock \textit{Phys. Rev. Lett.} \textbf{117}, 078301.

\bibitem[{Nobile(2004)}]{Nobile04}
Nobile, A., 2004.
\newblock On the posterior distribution of the number of components in a finite
  mixture.
\newblock \textit{The Annals of Statistics} \textbf{32}, 2044--2073.

\bibitem[{Nobile and Fearnside(2007)}]{NF07}
Nobile, A. and Fearnside, A., 2007.
\newblock Bayesian finite mixtures with an unknown number of components: The
  allocation sampler.
\newblock \textit{Stat. Comput.} \textbf{17}, 147--162.

\bibitem[{Nylund \textit{et~al.}(2007)Nylund, Asparouhov, and
  Muth{\'u}n}]{NAM07}
Nylund, K.~L., Asparouhov, T., and Muth{\'u}n, B.~O., 2007.
\newblock Deciding on the number of classes in latent class analysis and growth
  mixture modeling: A {M}onte {C}arlo simulation study.
\newblock \textit{Structural Equation Modeling} \textbf{14}, 535--569.

\bibitem[{Phillips and Smith(1996)}]{PS96}
Phillips, D.~B. and Smith, A. F.~M., 1996.
\newblock Bayesian model comparison via jump diffusions.
\newblock In W.~R. Gilks, S.~Richardson, and D.~J. Spiegelhalter (eds.),
  \textit{Markov Chain Monte Carlo in Practice}, pp. 215--239. Chapman and
  Hall, New York.

\bibitem[{Porteous \textit{et~al.}(2008)Porteous, Newman, Ihler, Asuncion,
  Smyth, and Welling}]{Porteous08}
Porteous, I., Newman, D., Ihler, A., Asuncion, A., Smyth, P., and Welling, M.,
  2008.
\newblock Fast collapsed {G}ibbs sampling for latent {D}irichlet allocation.
\newblock In \textit{Proceedings of the 14th {ACM} {SIGKDD} International
  Conference on Knowledge Discovery and Data Mining}, pp. 569--577. Association
  of Computing Machinery, New York.

\bibitem[{Raftery and Dean(2006)}]{RD06}
Raftery, A.~E. and Dean, N., 2006.
\newblock Variable selection for model-based clustering.
\newblock \textit{Journal of the American Statistical Association}
  \textbf{101}, 168--178.

\bibitem[{Richardson and Green(1997)}]{RG97}
Richardson, S. and Green, P.~J., 1997.
\newblock On {B}ayesian analysis of mixtures with an unknown number of
  components.
\newblock \textit{Journal of the Royal Statistical Society B} \textbf{59},
  731--792.

\bibitem[{Riolo and Newman(2020)}]{RN20}
Riolo, M.~A. and Newman, M. E.~J., 2020.
\newblock Consistency of community structure in complex networks.
\newblock \textit{Phys. Rev. E} \textbf{101}, 052306.

\bibitem[{Riyanto \textit{et~al.}(2022)Riyanto, Kuswanto, and Prastyo}]{RHP22}
Riyanto, A., Kuswanto, H., and Prastyo, D.~D., 2022.
\newblock Mutual information-based variable selection on latent class cluster
  analysis.
\newblock \textit{Symmetry} \textbf{14}, 908.

\bibitem[{Rousseau and Mengersen(2011)}]{RM11}
Rousseau, J. and Mengersen, K., 2011.
\newblock Asymptotic behaviour of the posterior distribution in overfitted
  mixture models.
\newblock \textit{Journal of the Royal Statistical Society B} \textbf{73},
  689--710.

\bibitem[{Steele \textit{et~al.}(2006)Steele, Raftery, and Emond}]{SRE06}
Steele, R.~J., Raftery, A.~E., and Emond, M.~J., 2006.
\newblock Computing normalizing constants for finite mixture models via
  incremental mixture importance sampling (imis).
\newblock \textit{Journal of Computational and Graphical Statistics}
  \textbf{15}, 712--734.

\bibitem[{Stephens(2000{\natexlab{a}})}]{Stephens00a}
Stephens, M., 2000{\natexlab{a}}.
\newblock Dealing with label switching in mixture models.
\newblock \textit{Journal of the Royal Statistical Society B} \textbf{62},
  795--809.

\bibitem[{Stephens(2000{\natexlab{b}})}]{Stephens00b}
Stephens, M., 2000{\natexlab{b}}.
\newblock Bayesian analysis of mixture models with an unknown number of
  components---an alternative to reversible jump methods.
\newblock \textit{Annals of Statistics} \textbf{28}, 40--74.

\bibitem[{Titterington \textit{et~al.}(1988)Titterington, Smith, and
  Makov}]{TSM85}
Titterington, D.~M., Smith, A. F.~M., and Makov, U.~E., 1988.
\newblock \textit{Statistical Analysis of Finite Mixture Distributions}.
\newblock John Wiley, New York.

\bibitem[{van Havre \textit{et~al.}(2015)van Havre, White, Rousseau, and
  Mengersen}]{VWRM15}
van Havre, Z., White, N., Rousseau, J., and Mengersen, K., 2015.
\newblock Overfitting {B}ayesian mixture models with an unknown number of
  components.
\newblock \textit{PLOS One} \textbf{10}, e0131739.

\bibitem[{Vega-Pons and Ruiz-Shulcloper(2011)}]{VR11}
Vega-Pons, S. and Ruiz-Shulcloper, J., 2011.
\newblock A survey of clustering ensemble algorithms.
\newblock \textit{International Journal of Pattern Recognition and Artificial
  Intelligence} \textbf{25}, 337--372.

\bibitem[{Walsh(2006)}]{Walsh06}
Walsh, C.~D., 2006.
\newblock Latent class analysis identification of syndromes in {A}lzheimer’s
  disease: A {B}ayesian approach.
\newblock \textit{Advances in Methodology and Statistics} \textbf{3}, 147--162.

\bibitem[{Wang \textit{et~al.}(1996)Wang, Puterman, Cockburn, and Le}]{WPCL96}
Wang, P., Puterman, M.~L., Cockburn, I., and Le, N., 1996.
\newblock Mixed {P}oisson regression models with covariate dependent rates.
\newblock \textit{Biometrics} \textbf{52}, 381--400.

\bibitem[{White \textit{et~al.}(2016)White, Wyse, and Murphy}]{WWM16}
White, A., Wyse, J., and Murphy, T.~B., 2016.
\newblock Bayesian variable selection for latent class analysis using a
  collapsed {G}ibbs sampler.
\newblock \textit{Stat. Comput.} \textbf{26}, 511--527.

\bibitem[{Zhou \textit{et~al.}(2023)Zhou, Dudeja, and Amini}]{ZDA23}
Zhou, Z., Dudeja, G., and Amini, A.~A., 2023.
\newblock Statistical guarantees for consensus clustering.
\newblock In \textit{Proceedings of the 11th International Conference on
  Learning Representations}. OpenReview.net.

\end{thebibliography}
\end{document}